\pdfoutput=1
\documentclass[10pt,conference]{IEEEtran}

\usepackage[T1]{fontenc}
\usepackage[utf8]{inputenc}
\usepackage{graphicx}
\usepackage{amsmath,amssymb}
\usepackage{mathtools}
\usepackage{amsthm}
\theoremstyle{definition}
\newtheorem{definition}{Definition}
\newtheorem{example}{Example}
\usepackage{booktabs}
\usepackage{url}
\usepackage{xurl}
\usepackage{pgfplots}
\usepackage{pgfplotstable}
\pgfplotsset{compat=1.18}
\usepackage{cite}

\usepackage[table]{xcolor}
\usepackage{hyperref}
\definecolor{linkblue}{RGB}{0,90,180}
\hypersetup{colorlinks=true,citecolor=linkblue,linkcolor=linkblue,urlcolor=linkblue}
\usepackage{enumitem}
\usepackage{xspace}
\usepackage{fontawesome5}
\usepackage{relsize}
\usepackage[most]{tcolorbox}
\usepackage{listings}
\usepackage{inconsolata}

\usepackage{adjustbox}
\usepackage{pdfrender}
\definecolor{bugboxbg}{RGB}{243,244,246}
\definecolor{bugboxrule}{RGB}{150,158,170}
\definecolor{codekw}{RGB}{43,76,140}
\definecolor{codestr}{RGB}{150,72,30}
\definecolor{codecom}{RGB}{122,130,142}
\lstdefinestyle{bugbox}{%
  basicstyle=\fontfamily{zi4}\selectfont\footnotesize,
  backgroundcolor=\color{bugboxbg},
  frame=leftline, framerule=2pt, rulecolor=\color{bugboxrule},
  xleftmargin=10pt, framexleftmargin=6pt, framexrightmargin=2pt, framesep=4pt,
  aboveskip=4pt, belowskip=4pt, lineskip=0.6pt,
  breaklines=false, columns=fullflexible, keepspaces=true,
  showstringspaces=false, language=Python,
  keywordstyle=\color{codekw}\bfseries,
  stringstyle=\color{codestr},
  commentstyle=\color{codecom}\itshape,
  morekeywords={await,async,self,None,True,False},
}
\lstnewenvironment{bugsnippet}{\lstset{style=bugbox}}{}
\newcommand{\blab}[1]{\mathmakebox[2.7em][l]{(\beta_{#1})}}
\newcommand{\bnolab}{\mathmakebox[2.7em][l]{}}
\newcommand{\fhead}[1]{\mathmakebox[3.8em][l]{\mathsf{fire}(#1)}}
\newcommand{\bgraph}[1]{\par\addvspace{3pt}\centerline{#1}\addvspace{2pt}\par}
\renewcommand{\footnoterule}{\kern-3pt\hrule width 0.4\columnwidth height 0.4pt\kern 2.4pt}
\usetikzlibrary{arrows.meta,positioning,calc,backgrounds,fit}
\definecolor{archSlate}{RGB}{100,116,139}
\definecolor{archStone}{RGB}{168,162,158}
\definecolor{archCoral}{RGB}{217,119,87}
\definecolor{rowshade}{gray}{0.92}
\tikzset{
  Pnode/.style={font=\scriptsize, draw=black!55, rounded corners=2pt,
    inner sep=2.5pt, minimum width=0.9cm, minimum height=0.42cm, align=center,
    fill=archSlate!12},
  Slot/.style={Pnode, fill=archStone!80},
  Ghost/.style={Pnode, draw=archCoral!75!black, dashed, fill=white,
    text=archCoral!78!black},
  Final/.style={circle, draw=black!65, fill=white, line width=0.6pt, inner sep=0pt,
    minimum size=8pt,
    path picture={\fill[black!78] (path picture bounding box.center) circle (1.6pt);}},
  Start/.style={circle, fill=black!75, inner sep=0pt, minimum size=5.5pt},
  Lab/.style={font=\tiny, text=black!72, align=center, inner sep=1pt},
  LabC/.style={font=\tiny, text=archCoral!78!black, align=center, inner sep=1pt},
  E/.style={-{Stealth[length=1.4mm,width=1.3mm]}, line width=0.7pt},
  F/.style={-{Stealth[length=1.4mm,width=1.3mm]}, line width=0.85pt,
    draw=archCoral!78!black},
  M/.style={-{Stealth[length=1.4mm,width=1.3mm]}, line width=0.75pt, dashed,
    draw=archCoral!78!black},
}

\newcommand{\tool}{{\textsc{Apsis}}\xspace}

\newcommand{\kw}[1]{\ensuremath{\mathsf{#1}}}
\newcommand{\FPR}{\ensuremath{\mathit{FPR}}\xspace}
\newcommand{\appref}[1]{\hyperref[#1]{Appendix~\ref*{#1}}}
\newcommand{\icode}[1]{%
  \ifmmode\text{\relsize{-0.5}\texttt{#1}}%
  \else\mbox{\relsize{-0.5}\texttt{#1}}\fi}

\newcommand{\subject}[1]{\par\smallskip\noindent\textbf{#1}\ }

\newcounter{takeaway}
\newcommand{\takeaways}[1]{
\vspace{0.5em}
\noindent
\begin{tcolorbox}[ enhanced,
    breakable,
    boxrule=1pt,
    arc=4pt,
    left=2pt,
    right=2pt,
    bottom=2pt,
    top=2pt,
    colback=gray!4,
    colframe=gray!1!black,
    drop shadow=black!50!white,
    rounded corners]
\noindent
\refstepcounter{takeaway}
\textbf{Takeaway \Roman{takeaway}.}
{#1}
\end{tcolorbox}
}

\makeatletter

\long\def\apsis@makecaption#1#2{%
  \ifx\@captype\@IEEEtablestring\else\@IEEEfigurecaptionsepspace\fi
  \setbox\@tempboxa\hbox{\normalfont\footnotesize {#1.}\nobreakspace\nobreakspace #2}%
  \ifdim \wd\@tempboxa >\hsize
    \setbox\@tempboxa\hbox{\normalfont\footnotesize {#1.}\nobreakspace\nobreakspace}%
    \parbox[t]{\hsize}{\normalfont\footnotesize\noindent\unhbox\@tempboxa#2}%
  \else
    \hbox to\hsize{\normalfont\footnotesize\hfil\box\@tempboxa\hfil}%
  \fi
  \ifx\@captype\@IEEEtablestring\@IEEEtablecaptionsepspace\fi}
\AtBeginDocument{\let\@makecaption\apsis@makecaption}
\makeatother

\begin{document}

\title{Automated Safety Analysis for Agentic Workflows}
\author{%
\IEEEauthorblockN{%
Yanju Chen\IEEEauthorrefmark{1},
Ying Li\IEEEauthorrefmark{2},
Bosi Zhang\IEEEauthorrefmark{3},
Hanzhi Liu\IEEEauthorrefmark{4},
Hongbo Wen\IEEEauthorrefmark{4},
Yufei Ding\IEEEauthorrefmark{1},
Yuan Tian\IEEEauthorrefmark{2} and
Yu Feng\IEEEauthorrefmark{4}}
\IEEEauthorblockA{\IEEEauthorrefmark{1}University of California, San Diego,
\{yanju, yufeiding\}@ucsd.edu}
\IEEEauthorblockA{\IEEEauthorrefmark{2}University of California, Los Angeles,
\{yinglee, yuant\}@ucla.edu}
\IEEEauthorblockA{\IEEEauthorrefmark{3}City University of Hong Kong,
bosi.zhang@my.cityu.edu.hk}
\IEEEauthorblockA{\IEEEauthorrefmark{4}University of California, Santa Barbara,
\{hanzhi, hongbowen\}@ucsb.edu, yufeng@cs.ucsb.edu}}

\maketitle

\makeatletter
\def\ps@extended{%
  \def\@oddhead{\hbox{}\normalfont\footnotesize Automated Safety Analysis for Agentic Workflows (Extended Version)\hfil\thepage}%
  \let\@evenhead\@oddhead
  \let\@oddfoot\@empty\let\@evenfoot\@empty}
\def\ps@extendedfirst{%
  \def\@oddhead{\hbox{}\normalfont\footnotesize Extended Version of the Paper Accepted at ICSE 2027, with Appendices\hfil\thepage}%
  \let\@evenhead\@oddhead
  \def\@oddfoot{\parbox[b]{\textwidth}{\normalfont\scriptsize \copyright~2027 IEEE. Personal use of this material is permitted. Permission from IEEE must be obtained for all other uses, in any current or future media, including reprinting/republishing this material for advertising or promotional purposes, creating new collective works, for resale or redistribution to servers or lists, or reuse of any copyrighted component of this work in other works.}}%
  \let\@evenfoot\@oddfoot}
\makeatother
\pagestyle{extended}
\thispagestyle{extendedfirst}

\begin{abstract}
LLM applications are increasingly built as \emph{agentic workflows}: programs
that combine model calls, external tools, and control logic to act autonomously
on a user's behalf. Because these workflows execute real-world actions, logical
flaws in how a workflow is wired can cause severe harm in deployed systems.
Unlike prompt injections or malformed API calls, these \emph{workflow bugs}
stem from defects in the workflow's design, yet no existing tool can statically
validate workflow logic across frameworks prior to deployment. We present
\tool, a static safety analyzer that detects workflow bugs before deployment.
To operate across frameworks, \tool lifts framework-specific code into a common
intermediate representation (IR). This IR models a workflow as a semantic graph
of steps, data dependencies, and shared state, while annotating each component
with behavioral properties such as whether an input is mandatory or how a
shared channel merges concurrent writes. Building on this representation, we
introduce a bug description language that formalizes recurring classes of
agentic failures as declarative patterns over the IR. By matching these
patterns against the IR, \tool detects structural and behavioral
flaws and pinpoints the exact step responsible for each bug. To evaluate \tool,
we construct, to our knowledge, the first corpus of workflow bugs, distilled from real-world reports across 13 widely used frameworks. On a reproducible benchmark suite drawn from this corpus, \tool detects
workflow bugs with 91\% precision and 84\% recall, and uncovers \textbf{36}
previously unknown bugs in real-world workflows.
\end{abstract}


\section{Introduction}
\label{sec:intro}

LLM applications are evolving from simple chatbots into \emph{agentic
workflows}: programs that chain together model calls, external tools, and
control logic to autonomously execute multi-step tasks~\cite{yao2023react}.
Developers increasingly build these workflows using orchestration frameworks
such as LangGraph~\cite{langgraph}, Microsoft's Agent
Framework~\cite{agentframework}, and Haystack~\cite{haystack}, among
others~\cite{wu2023autogen}, by composing agents, tools, shared state, and
branching control flow. However, because an agentic workflow acts in the real
world by calling APIs, executing code, and modifying data, a flaw in its logic
can trigger destructive actions rather than merely producing an incorrect text
response.

These logical flaws have already caused severe real-world harm. For example,
given a deployment task it could not complete, the coding agent
Devin~\cite{devin} lacked a halting condition and looped for over a day,
draining compute and API budgets without producing a
result~\cite{husain2025devin,claburn2025devin}. In another incident, a Cursor
agent attempting to resolve a configuration error wiped a company's entire
production database and every backup in seconds because the workflow ran with
an over-privileged credential and no human confirmation step; the data was
recoverable only from an internal snapshot that the agent never knew existed~\cite{pocketos,zenity2026pocketos}. Crucially, neither failure was caused
by a prompt injection~\cite{greshake2023injection} or a malformed API call.
Instead, both incidents resulted from flaws in the workflow's own control and
coordination logic, which we define as \emph{workflow bugs}.

Despite the high stakes, to our knowledge no existing tool validates an agentic workflow's logic prior to execution; adjacent lines of research target orthogonal
problems. For instance, security research on agent frameworks hardens the
underlying execution engines~\cite{checkpoint2026langgraph}, but provides no
guarantees for the developer-defined workflows built on top of those engines.
Meanwhile, empirical studies catalog how deployed agents
fail~\cite{cemri2025mast}, characterizing failures after execution rather than
preventing them beforehand. General-purpose static analyzers and
linters~\cite{codeql,infer} inspect host-language source code, where a workflow
is constructed through ordinary function calls rather than represented
explicitly as a connected, stateful graph. Finally, while some frameworks
provide built-in validation (such as type-checked component connections in
Haystack~\cite{haystack}), these checks catch only basic structural errors and
are confined to a single framework. Consequently, developers lack a
cross-framework static analyzer capable of reasoning about a workflow's logic
before deployment.

Building such an analyzer poses three key challenges. First, the analysis must
be \emph{static}: workflow bugs must be detected before execution, because
consequences such as data loss or budget exhaustion are irreversible. Second,
the analysis must be \emph{semantic}: many workflow bugs arise not from graph
topology alone, but from step-level execution semantics, such as a lost
concurrent write~\cite{ghLanggraph1406}, a mis-ordered
confirmation~\cite{ghLanggraphjs792,ghDeepagents554}, or a capability
over-reach~\cite{ghAutogen960}. Third, the analysis must be \emph{general}: an
analyzer tied to one framework cannot be reused when the same workflow logic is
expressed in another, even when the underlying workflows are conceptually
identical.

To address these challenges, we present \tool, a static safety analyzer that
detects logical flaws in agentic workflows before deployment. To operate across
frameworks, \tool lifts framework-specific code into a common intermediate
representation (IR). In particular, the IR models a workflow as a graph whose
\emph{nodes} represent workflow steps (agents, tool calls, routers, and
subflows), whose \emph{edges} capture control and data dependencies, and whose
\emph{state} represents shared data channels. To enable semantic reasoning
beyond graph topology, the IR annotates each node and edge with behavioral labels,
such as whether an input is mandatory, what role a step carries, or how a
shared channel merges concurrent writes. Because this IR is framework-neutral,
a single suite of checks can analyze workflows across frameworks, and
supporting a new framework requires only a front-end mapping to the IR.

Over this representation, \tool runs structural checks to catch topological
flaws (e.g., an inescapable loop~\cite{ghLanggraph96,ghAutogen5248} or an
invalid router target~\cite{ghAutogen781}) alongside behavioral checks to catch
semantic errors (e.g., a lost concurrent write~\cite{ghLanggraph1406} or a
mis-ordered confirmation~\cite{ghLanggraphjs792,ghDeepagents554}). To express
these checks systematically, we design a domain-specific \emph{bug description
language} (BDL) that formalizes each workflow bug as a declarative pattern over
the nodes, edges, state, and behavioral labels of the IR. By matching these
patterns against a workflow's IR, \tool detects violations statically and
pinpoints the exact step or connection responsible for each bug. Moreover,
decoupling bug definitions from the analysis engine lets developers encode new
agentic failure modes as declarative rules.

To characterize workflow bugs and evaluate \tool, we collected 861 issue
reports across 13 widely used open-source orchestration frameworks. From a
representative sample, we derive a \emph{taxonomy} of workflow bugs and
formalize a pattern for each category. From the full set, we construct, to our
knowledge, the first labeled \emph{corpus} of workflow bugs, and draw from this
corpus a reproducible \emph{benchmark suite} of $135$ buggy workflows and their fixes.
On this suite, \tool detects workflow bugs with $91\%$ precision and $84\%$
recall ($F_1 = 0.87$) in $0.7$\,s per workflow, outperforming the strongest
baseline by $0.18$~$F_1$. In particular, \tool's behavioral checks detect $46$
bugs that structural checks alone miss. Applied to real-world workflows, \tool uncovers \textbf{36} previously unknown bugs.

\subject{Contributions.}
We make the following contributions:
\begin{itemize}[noitemsep, topsep=1pt, leftmargin=1.5em]
\item \textbf{A taxonomy and the first corpus of workflow bugs.} From 861
      real-world reports across 13 widely used frameworks, we derive a taxonomy
      and the first labeled corpus of workflow bugs, and draw from this corpus
      a reproducible suite of $135$ benchmarks.
\item \textbf{A semantic intermediate representation.} We design a
      framework-neutral IR that captures both the topology and behavioral
      semantics of agentic workflows, enabling a single analysis to reason
      across orchestration frameworks.
\item \textbf{A formal bug description language.} We introduce a
      domain-specific language that formalizes workflow bugs as declarative
      graph patterns over the IR, decoupling bug definitions from the analysis
      engine so that new failure modes can be encoded as executable rules.
\item \textbf{The \tool analyzer.} We build \tool, an end-to-end static
      analyzer that lifts workflows into the IR, matches formal bug patterns
      against this representation, and pinpoints the exact step responsible for
      each flaw before execution.
\item \textbf{Evaluation.} On our benchmarks, \tool achieves $91\%$ precision and
      $84\%$ recall ($F_1 = 0.87$), outperforming the strongest baseline by
      $0.18$~$F_1$. An ablation study shows that behavioral checks catch $46$
      bugs missed by structural checks alone, and \tool uncovers \textbf{36}
      previously unknown bugs in real-world workflows.
\end{itemize}

\section{Background and Overview}
\label{sec:background}

\subsection{Agentic Workflows as Graphs}
\label{sec:background:graphs}

\begin{figure}[t]
\centering
\definecolor{wfAccent}{RGB}{100,116,139}
\definecolor{wfFill}{RGB}{238,240,242}
\definecolor{aiZone}{RGB}{240,241,244}
\definecolor{aiEdge}{RGB}{188,193,203}
\definecolor{inkDark}{RGB}{30,41,59}
\adjustbox{max width=\columnwidth}{%
\begin{tikzpicture}[
  font=\sffamily, >={Stealth[length=2.0mm,width=1.9mm]},
  fwd/.style={->,draw=archSlate,line width=0.9pt},
  ret/.style={->,draw=archSlate!70,line width=0.8pt,densely dashed},
  drive/.style={->,draw=wfAccent,line width=1.2pt},
  inner/.style={->,draw=archSlate,line width=0.8pt},
  lbl/.style={font=\sffamily\scriptsize,text=archSlate,inner sep=1.5pt},
  ctrl/.style={rounded corners=2pt,draw=archSlate,line width=0.8pt,fill=white,
    inner xsep=4pt,inner ysep=3.5pt,align=center,text=inkDark,font=\sffamily\small},
  sub/.style={rounded corners=2pt,draw=archSlate!70,line width=0.8pt,fill=white,
    inner xsep=4pt,inner ysep=3pt,align=center,text=inkDark,font=\sffamily\small},
]
  \node[ctrl] (route) {\faRandom~route};
  \node[sub,above right=5mm and 9mm of route.east,anchor=west] (sa1) {{\color{wfAccent}\faRobot}~~sub-agent};
  \node[sub,below right=5mm and 9mm of route.east,anchor=west] (sa2) {{\color{wfAccent}\faRobot}~~sub-agent};
  \node[ctrl,right=9mm of sa1.east |- route] (merge) {\faObjectGroup~merge};
  \begin{scope}[on background layer]
    \node[draw=aiEdge,dash pattern=on 1.3pt off 1.3pt,rounded corners=3pt,fill=aiZone,
          fit=(sa1)(sa2),inner sep=3pt] (aigrp) {};
  \end{scope}
  \draw[inner] (route.east) -- (sa1.west);
  \draw[inner] (route.east) -- (sa2.west);
  \draw[inner] (sa1.east) -- (merge.west);
  \draw[inner] (sa2.east) -- (merge.west);
  \coordinate (tp) at ($(aigrp.north)+(0,6.5mm)$);
  \begin{scope}[on background layer]
    \node[draw=wfAccent,line width=1.4pt,rounded corners=5pt,fill=wfFill,
          fit=(route)(aigrp)(merge)(tp),inner xsep=6pt,inner ysep=4pt] (wf) {};
  \end{scope}
  \node[anchor=north,fill=wfAccent,text=white,font=\sffamily\bfseries\small,
        rounded corners=3pt,inner xsep=5pt,inner ysep=2.5pt]
        at ([yshift=-2.0mm]wf.north) {\faProjectDiagram~~Agentic Workflow};
  \node[rounded corners=3pt,draw=archSlate,line width=0.9pt,fill=white,
        minimum width=15mm,minimum height=14mm,align=center,text=inkDark,
        left=16mm of wf.west] (main) {{\large\color{wfAccent}\faRobot}\\[1pt]\small main\\[-1pt]\small agent};
  \node[rounded corners=3pt,draw=archSlate,line width=0.9pt,fill=white,
        minimum width=12mm,minimum height=14mm,align=center,text=inkDark,
        left=14mm of main] (user) {{\large\color{archSlate}\faUser}\\[1pt]\small user};
  \draw[fwd]   ([yshift=2.6mm]user.east) -- node[lbl,above]{task} ([yshift=2.6mm]main.west);
  \draw[ret]   ([yshift=-2.6mm]main.west) -- node[lbl,below]{answer} ([yshift=-2.6mm]user.east);
  \draw[drive] ([yshift=2.6mm]main.east) -- node[lbl,above,text=wfAccent]{drives} ([yshift=2.6mm]wf.west);
  \draw[ret]   ([yshift=-2.6mm]wf.west) -- node[lbl,below]{results} ([yshift=-2.6mm]main.east);
\end{tikzpicture}%
}
\caption{An agentic workflow's niche. Within the user-agent loop, a main agent does not
improvise: it drives a fixed workflow that routes a task to sub-agents and merges
their results before answering the user.}
\label{fig:niche}
\end{figure}

\autoref{fig:niche} illustrates the role of an \emph{agentic workflow} in an LLM
application. To handle a user's task, a main agent coordinates specialist
sub-agents. If the main agent organizes this work purely through free-form
prompting, execution is nondeterministic: it may loop, skip a required step, or
act before checking a precondition. To enforce predictability, the agent instead drives a \emph{workflow}: a fixed graph that connects nondeterministic
sub-agents into a deterministic control structure. As shown in
\autoref{fig:niche}, the workflow routes the task, merges results, and gates
irreversible actions before returning the outcome to the user-agent loop.

Although frameworks expose different interfaces (such as a Python builder in
LangGraph~\cite{langgraph}, a chainable API in Microsoft's Agent
Framework~\cite{agentframework}, a component pipeline in
Haystack~\cite{haystack}, or visual-editor JSON in Flowise~\cite{flowise} and
Langflow~\cite{langflow}), they all construct the same underlying graph. Its
\emph{nodes} are workflow steps (agents, tools, routers that select which branch
runs next, loops, subflows that embed a nested workflow, and human checkpoints
that pause for approval), its \emph{typed edges} carry data and hand off
control, and its \emph{channels} are named slots of shared state governed by a
reducer that merges concurrent writes. Because this graph is fixed when the
developer wires the workflow, its structure can be extracted directly from the
source code.

\begin{figure}[t]
\begin{lstlisting}[style=bugbox]
g.add_node("research", research)
g.add_node("analyze", analyze)
g.add_node("verify", verify)
g.add_edge("research", "analyze")
g.add_edge("analyze", "verify")
g.add_edge("verify", "analyze")   # always loops back
g.compile()   # analyze<->verify never reaches END
\end{lstlisting}
\caption{Running-example source (LangGraph): \icode{verify} always routes back to
\icode{analyze}, so the \icode{analyze}--\icode{verify} cycle is never wired onward to a
terminal and never halts, re-invoking both agents without end so the run burns budget and returns nothing.}
\label{fig:running}
\end{figure}

\begin{figure}[t]
\centering
\adjustbox{max width=\columnwidth, center}{%
\begin{tikzpicture}
  \node[Pnode] (res) at (0,0)    {$\icode{research} \dblcolon \textbf{agent}$};
  \node[Pnode] (ana) at (3.1,0)  {$\icode{analyze} \dblcolon \textbf{agent}$};
  \node[Pnode] (ver) at (6.2,0)  {$\icode{verify} \dblcolon \textbf{agent}$};
  \node[Final] (end) at (8.0,0) {};
  \draw[E] (res) -- (ana) node[Lab,above,midway,inner sep=1.5pt]{(\kw{control})};
  \draw[E] (ana) to[bend left=20] node[Lab,above,midway,inner sep=1.5pt]{(\kw{control})} (ver);
  \draw[E] (ver) to[bend left=20] node[Lab,below,midway,inner sep=1.5pt]{(\kw{control})} (ana);
  \draw[M] (ver) -- (end) node[LabC,above,midway,inner sep=1.5pt]{(\kw{control})};
\end{tikzpicture}%
}
\caption{The same workflow lifted into the IR: slate boxes are steps
(\texttt{name\,::\,Type}), arrows are control edges, the ringed dot is the terminal that the
\icode{analyze}--\icode{verify} cycle never reaches, and the dashed coral arrow is the absent
exit edge that is the defect.}
\label{fig:ir}
\end{figure}

\subsection{Running Example}
\label{sec:background:example}

In a real-world incident, a single workflow run accumulated a \$47{,}000 API
bill before anyone noticed, nearly a thousand times the \$50 a bounded run should have stayed under~\cite{anhaia2025loop}, due to a runaway multi-agent loop (our running
example). \autoref{fig:running} shows its LangGraph source: a research pipeline where \icode{research} passes control to
\icode{analyze}, which routes to \icode{verify}. However, \icode{verify} unconditionally routes back to
\icode{analyze}, so the \icode{analyze}--\icode{verify} cycle is never wired to
a terminal. Read in isolation, each step is well-formed; yet once control enters
the cycle it never exits, re-invoking both agents indefinitely.

Existing developer tools fail to catch this defect before deployment. The
framework's validator type-checks connections and compiles the graph without
error; a general static analyzer sees only the host program's \icode{add\_node}
and \icode{add\_edge} calls rather than the workflow graph they build; and a
runtime guardrail or LLM judge detects the runaway spend only while the loop is
running in production. Catching this bug beforehand requires reasoning about how
the wired steps interact under the engine's execution semantics.

\subject{Insight.}
Once a workflow's wiring and its framework's implicit execution semantics are
made explicit in a graph representation, the defect becomes a violated static
property of that graph rather than a runtime event. Once the source is lifted into the \tool IR (\autoref{fig:ir}), the \icode{analyze}--\icode{verify} cycle reaches
no terminal, so a declarative detector matches the unbounded cycle and
pinpoints the missing exit before execution.

\subject{Our Framework.}
\autoref{fig:overview} generalizes this process into \tool's three-stage
pipeline. A framework-specific front-end lifts a workflow into the common \tool
IR; the bug analysis engine matches the pattern catalog over the IR; and each
match produces a bug report pinpointing the responsible step. Because the IR and
analysis engine are shared, supporting a new framework requires only one
front-end.

\begin{figure}[t]
\centering
\definecolor{toolAccent}{RGB}{100,116,139}
\definecolor{toolFill}{RGB}{238,240,242}
\definecolor{ioFill}{RGB}{240,241,244}
\definecolor{bugRed}{RGB}{183,86,72}
\definecolor{inkDark}{RGB}{30,41,59}
\adjustbox{max width=\columnwidth}{%
\begin{tikzpicture}[
  font=\sffamily, >={Stealth[length=2.2mm,width=2.0mm]},
  flow/.style={->,draw=archSlate,line width=1.0pt},
  inner/.style={->,draw=toolAccent,line width=0.9pt},
  io/.style={rounded corners=2pt,draw=archSlate!70,line width=0.8pt,fill=ioFill,
    align=center,text=inkDark,inner xsep=5pt,inner ysep=5pt,font=\sffamily\small,minimum height=9mm},
  stage/.style={rounded corners=2pt,draw=toolAccent!70,line width=0.8pt,fill=white,
    align=center,text=inkDark,inner xsep=5pt,inner ysep=5pt,font=\sffamily\small,minimum height=9mm},
]
  \node[stage] (fe) {framework\\front-end};
  \node[stage,right=9mm of fe] (ir) {\tool\ IR};
  \node[stage,right=9mm of ir] (an) {bug\\analysis};
  \draw[inner] (fe) -- (ir);
  \draw[inner] (ir) -- (an);
  \coordinate (tp) at ($(ir.north)+(0,6mm)$);
  \begin{scope}[on background layer]
    \node[draw=toolAccent,line width=1.4pt,rounded corners=5pt,fill=toolFill,
          fit=(fe)(an)(tp),inner xsep=7pt,inner ysep=5pt] (apsis) {};
  \end{scope}
  \begin{scope}[on background layer]
    \fill[white,draw=archSlate!55,line width=0.6pt,rounded corners=2pt] ([shift={(1.4mm,1.4mm)}]fe.south west) rectangle ([shift={(1.4mm,1.4mm)}]fe.north east);
    \fill[white,draw=archSlate!55,line width=0.6pt,rounded corners=2pt] ([shift={(0.7mm,0.7mm)}]fe.south west) rectangle ([shift={(0.7mm,0.7mm)}]fe.north east);
  \end{scope}
  \node[anchor=north,fill=toolAccent,text=white,font=\sffamily\bfseries\small,
        rounded corners=3pt,inner xsep=5pt,inner ysep=2.5pt]
        at ([yshift=-2.0mm]apsis.north) {\textsc{Apsis} (Our Tool)};
  \node[io,left=11mm of apsis.west] (wf) {{\color{archSlate}\faProjectDiagram}~~workflow};
  \begin{scope}[on background layer]
    \fill[ioFill,draw=archSlate!55,line width=0.6pt,rounded corners=2pt] ([shift={(1.4mm,1.4mm)}]wf.south west) rectangle ([shift={(1.4mm,1.4mm)}]wf.north east);
    \fill[ioFill,draw=archSlate!55,line width=0.6pt,rounded corners=2pt] ([shift={(0.7mm,0.7mm)}]wf.south west) rectangle ([shift={(0.7mm,0.7mm)}]wf.north east);
  \end{scope}
  \node[io,right=11mm of apsis.east] (br) {{\color{bugRed}\faBug}~~bug report};
  \draw[flow] (wf.east) -- (apsis.west);
  \draw[flow] (apsis.east) -- (br.west);
\end{tikzpicture}%
}
\caption{\tool's pipeline: from a workflow to a bug report.}
\label{fig:overview}
\end{figure}

\subsection{Scope}
\label{sec:background:scope}

\tool's safety analysis targets workflow bugs like the runaway loop above. Such a bug lives in the workflow's wired logic, is detectable before execution from the workflow as written (its graph, its shared state, and the behavioral labels on its steps), and ends in an unsafe \emph{action}.
Examples include budget exhaustion from unbounded cycles, silently
corrupted results from overwritten channels, and irreversible effects from
unguarded steps. \tool analyzes a workflow under its framework's execution
semantics, assuming only that the workflow is available for static inspection.

\section{A Taxonomy of Workflow Bugs}
\label{sec:taxonomy}

To ground \tool in the failures developers encounter in practice, we studied bug
reports from open-source orchestration frameworks. The study yields a labeled
bug corpus and a taxonomy of recurring bug categories, each anchored in a
classical analysis primitive.

\subsection{Study Methodology}
\label{sec:taxonomy:method}

We surveyed the public issue trackers of $13$ widely used open-source
orchestration frameworks, covering both code-first libraries and visual builders,
and gathered the $861$ reports listed in \autoref{tab:corpus}. Against a fixed
criterion, we identified \emph{workflow bugs}: defects whose root cause lies in
the workflow's wired logic and leaves a static graph signature separating the
buggy artifact from its fix. Excluding configuration, environment, and model-quality reports yielded $783$ workflow bugs.

We distill these categories from a design set of $31$ reports, chosen for the popularity and
diversity of their discussion: each category is organized around a classical analysis primitive,
illustrated by one or two bugs, and the detectors that follow are written against those primitives.

\begin{table}[t]
\centering
\caption{Per-framework reports surveyed and the workflow bugs among them, split into \emph{structural} (topology) and \emph{behavioral} (step-level) bugs; \emph{Stars} is GitHub star count, a coarse adoption proxy.}
\label{tab:corpus}
\footnotesize
\setlength{\tabcolsep}{3pt}
\rowcolors{2}{white}{rowshade}
\begin{tabular}{l r r r r r}
\toprule
Framework & Reports & Workflow & Struct. & Behav. & Stars \\
\midrule
Agent Framework~\cite{agentframework} & 249 & 214 & 16\% & 84\% & 11.5k \\
Flowise~\cite{flowise}      & 168 & 154 & 25\% & 75\% & 53.8k \\
LangGraph~\cite{langgraph}  & 150 & 136 & 21\% & 79\% & 35.3k \\
AutoGen~\cite{wu2023autogen}&  73 &  69 & 38\% & 62\% & 59.1k \\
Langflow~\cite{langflow}    &  60 &  59 & 36\% & 64\% & 149.9k \\
Haystack~\cite{haystack}    &  44 &  43 & 49\% & 51\% & 25.6k \\
\midrule
Others (7)~\cite{deepagents,llamaindex,langgraphjs,crewai,pydanticai,opendeepresearch,langchain}
                          & 117 & 108 & --- & --- & --- \\
\midrule
\textbf{Total}            & \textbf{861} & \textbf{783}
                          & \textbf{$\approx$25\%} & \textbf{$\approx$75\%} & \\
\bottomrule
\end{tabular}
\end{table}

\subsection{Recurring Bug Categories}
\label{sec:taxonomy:families}

\autoref{tab:taxonomy} summarizes the six recurring bug categories, pairing each
with its analysis primitive and static signature. The categories span a spectrum
from \emph{structural} to \emph{behavioral}: at the structural end, a defect is
visible in the topology alone (e.g., a route to a nonexistent node or a cycle
with no exit); at the behavioral end, the graph is well-formed and the fault lies in a step's declared behavior. Roughly a quarter of bugs are structural
and three quarters are behavioral. We describe each category below.

\begin{table*}[t]
\centering
\caption{Recurring categories of workflow bugs observed in our study, each pairing an
analysis primitive with the workflow property it protects.}
\label{tab:taxonomy}
\footnotesize
\renewcommand{\arraystretch}{0.95}
\rowcolors{2}{white}{rowshade}
\begin{tabular}{p{4.0cm} p{2.5cm} p{6.8cm} p{3.0cm}}
\toprule
\textbf{Category} & \textbf{Primitive} & \textbf{Static signature (buggy vs.\ fixed)} & \textbf{Repr.\ report} \\
\midrule
Stalled \& Runaway Workflows & reachability &
A required step never becomes ready, or a cycle reaches no terminal,
stalling the workflow. & LangGraph loop~\cite{ghLanggraph96} \\Broken Data Handoffs & type checking &
A producer's output type is not assignable to the consumer, or a step
writes an undeclared channel. & Agent Framework~\cite{ghAgentframework2181} \\Misrouted Dispatch & name resolution &
A route names no node, mutually exclusive branches converge, or a branch
reaches no terminal. & LangGraph router~\cite{ghLanggraph5556} \\Corrupted Shared State & non-interference &
Concurrent writes go unreconciled, a read precedes its write, or two steps
alias one instance. & LangGraph state~\cite{ghLanggraph1406} \\Unwired Subflows & inter-procedural &
A subflow's required input port is left unbound across the call boundary,
starving the nested graph. & Agent Framework~\cite{ghAgentframework3766} \\Role \& Interrupt Violations & role / protocol &
A binding exceeds a step's role capability, or a human-in-the-loop
interrupt is mis-ordered. & LangGraph.js~\cite{ghLanggraphjs792} \\
\bottomrule
\end{tabular}
\end{table*}

\subject{Stalled and runaway workflows.}
This category adapts reachability and well-foundedness: every required step should eventually become ready, and the workflow should halt. In the LangGraph
snippet below, \icode{orphan} is registered via \icode{add\_node} but never
connected, yet \icode{compile()} silently accepts the unreachable
step~\cite{ghLanggraph6735}; the dual failure is a loop that never halts, as in
the Devin incident~\cite{husain2025devin}.
\begin{bugsnippet}
g = StateGraph(State)
g.add_node("orphan", orphan_node)  # never linked
return g.compile()    # compiles silently
\end{bugsnippet}

\subject{Broken data handoffs.}
This category adapts type checking to data edges and shared channels: every consumer
should receive a value of a type it accepts, and distinct outputs should not
collide on one channel. In the Agent Framework
snippet below, a handler emits a \icode{JailbreakDecision} that no
downstream handler accepts, creating a statically dead
edge~\cite{ghAgentframework2181}. This is our most structural category.
\begin{bugsnippet}
@handler
async def run(self, ...) -> None:
    # no downstream handler accepts this type
    await ctx.send_message(JailbreakDecision(...))
\end{bugsnippet}

\subject{Misrouted dispatch.}
This category adapts control-flow and name resolution to routers and conditional
edges: every declared route should resolve to a reachable, completable target. It is violated when a route references a nonexistent node (as in the
LangGraph snippet below~\cite{ghLanggraph5556}), when mutually exclusive branches converge on the same successor, or when a branch dead-ends without
reaching any terminal node.
\begin{bugsnippet}
def researcher_llm_call(state) -> Command:
    # typo: registered as "researcher_tools"
    return Command(goto="researcher_tool_node")
\end{bugsnippet}

\subject{Corrupted shared state.}
This category adapts non-interference (no concurrent step corrupts another's state) to shared channels:
concurrent writes should be reconciled, reads should see the writes they depend on,
and no two steps should alias one mutable instance. In the LangGraph snippet
below, parallel workers concurrently write to the \icode{result} channel with no
reducer, so one write silently overwrites the other~\cite{ghLanggraph1406}. The
defect is behavioral: topology shows two writers, while the lost update depends
on the channel's declared reducer.
\begin{bugsnippet}
class State(TypedDict):
    result: str    # LastValue: clobbers, no reducer
def worker(state) -> dict:
    return {"result": ...}  # concurrent writes
\end{bugsnippet}

\subject{Unwired subflows.}
This category adapts inter-procedural binding to subflow references across
graphs: every input a subflow declares should be bound at each call site. In the Agent Framework snippet below, the parent wires the
sub-workflow host \icode{\_host} downstream but leaves its request port
unbound, starving the nested graph~\cite{ghAgentframework3766}. The defect is
invisible in either graph alone, surfacing only once composed.
\begin{bugsnippet}
workflow = (WorkflowBuilder(start=_start)
    .add_edge(_main_agent, _host)
    .add_edge(_host, _end)  # request port unbound
    .build())
\end{bugsnippet}

\subject{Role and interrupt violations.}
This category adapts role and protocol well-formedness and is the most purely
behavioral, querying step labels rather than topology. A step should act
only within its role's capability. In the AutoGen snippet below, a
\icode{GroupChatManager} meant solely to coordinate turns is assigned its own
tools (\icode{tool\_cfg}), exceeding its role~\cite{ghAutogen960}; the category
also covers human-in-the-loop interrupts that fire in the wrong order or before
any agent produces work to confirm.
\begin{bugsnippet}
manager = GroupChatManager(
    groupchat=groupchat, ...)
manager.llm_config = tool_cfg  # manager gets tools
\end{bugsnippet}

\subsection{Key Lessons}
\label{sec:taxonomy:lessons}

Our study yields two key lessons that motivate \tool's design.
\textbf{Behavioral semantics beyond topology:} roughly three quarters of
workflow bugs are invisible in the topology alone and instead depend on a step's
declared behavior (whether an input is mandatory, how a channel reconciles
concurrent writes, or what capabilities a role permits), so an analyzer over bare nodes and edges is bounded by the structural quarter.
\textbf{Cross-framework equivalence:} a missing state reducer in LangGraph and
an early-firing join in Flowise represent the same non-interference violation,
and an unresolved route manifests identically whether written as a Python string
or a node id in an exported JSON flow. A framework-neutral, behaviorally
annotated representation thus allows a single suite of declarative patterns to
detect these bugs across frameworks.

\section{The Workflow IR}
\label{sec:ir}

A workflow bug is a violated property of a workflow's orchestration logic. Because developers express this logic via framework-specific APIs while the runtime executes it under implicit firing, merging, and routing rules, defects arise from the interaction of graph structure and execution semantics. To analyze workflows uniformly across frameworks, \tool lifts each workflow into an intermediate representation (IR) pairing the wired graph with explicit behavioral annotations, so each bug becomes a declarative property of the graph under its semantics.

\subsection{Preliminaries}
\label{sec:ir:abstraction}

Our analysis operates over a single workflow, which pairs a typed orchestration graph with the behavioral annotations and execution model that govern its runtime execution.

\begin{definition}[Workflow]
\label{def:workflow}
A workflow is a tuple
\[
  W = (V,\ E,\ \Sigma,\ \kappa,\ \lambda),
\]
where
\begin{itemize}[noitemsep, topsep=1pt, leftmargin=1.4em]
\item $V$ is a set of \emph{steps} (nodes), each with input and output ports;
\item $E \subseteq V \times V \times \Delta$ is a set of \emph{typed edges} of kind $\delta \in \Delta = \{\kw{data}, \kw{control}, \kw{cond}, \kw{tool}, \kw{handoff}, \kw{subflow}\}$: a \kw{data} edge passes a value from producer to consumer, a \kw{tool} edge binds a tool to its caller, a \kw{handoff} edge transfers control between agents, and a \kw{cond} edge carries a branch label, written $\mathsf{branch}(u,v,\ell)$;
\item $\Sigma$ is the set of declared \emph{channels} (named shared state); each
  $c \in \Sigma$ carries a reducer $\rho$, written $\mathsf{reducer}(c,\rho)$, resolving concurrent writes via $\kw{lastValue}$ (last-writer-wins) or a custom merge function;
\item $\kappa : V \to K$ assigns each step a \emph{kind} from $K = \{\kw{agent},
  \kw{tool}, \kw{router}, \kw{loop}, \kw{human}, \kw{subflow},
  \kw{start}, \kw{end}\}$,
  where \kw{human} is a human-in-the-loop step, \kw{loop} is a structural loop node in a visual builder, and \kw{subflow} is a step that embeds another workflow;
\item $\lambda : V \cup E \to 2^{\Lambda}$ assigns \emph{behavioral labels} from $\Lambda$ capturing semantics beyond topology, such as mandatory or defaulted inputs, declared routing targets, and step roles.
\end{itemize}
Each workflow also carries an \emph{execution model} from $\{\kw{dataflow}, \kw{bsp}, \kw{groupChat}, \kw{visual}\}$, set by the front-end and queried via $\mathsf{execModel}$, fixing how the engine schedules steps and merges state: $\kw{dataflow}$ fires steps as inputs arrive, $\kw{bsp}$ runs bulk-synchronous supersteps separated by barriers, $\kw{groupChat}$ routes messages among agents, and $\kw{visual}$ runs a flow exported from a visual builder.
\end{definition}

\begin{example}[A workflow lifted into the IR]\label{ex:workflow}
Slate boxes denote steps, the stone box is the shared channel \icode{reply}, and the ringed dot is the terminal; each arrow is labeled with its IR fact.
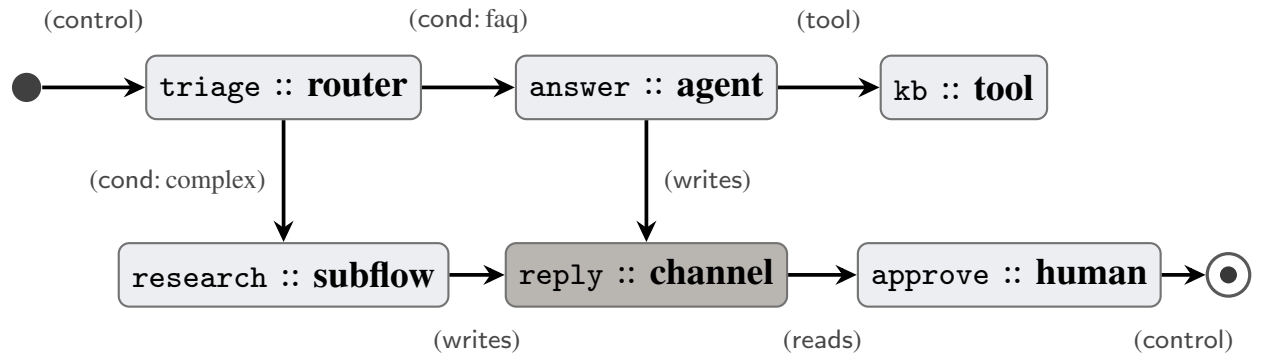
\begin{center}
\adjustbox{max width=\columnwidth}{%
\begin{tikzpicture}
  \node[Start] (s)   at (0,0.62)    {};
  \node[Pnode] (tri) at (1.7,0.62)  {$\icode{triage} \dblcolon \textbf{router}$};
  \node[Pnode] (ans) at (4.1,0.62)  {$\icode{answer} \dblcolon \textbf{agent}$};
  \node[Pnode] (kb)  at (6.2,0.62)  {$\icode{kb} \dblcolon \textbf{tool}$};
  \node[Pnode] (res) at (1.7,-0.62)  {$\icode{research} \dblcolon \textbf{subflow}$};
  \node[Slot]  (rep) at (4.1,-0.62)  {$\icode{reply} \dblcolon \textbf{channel}$};
  \node[Pnode] (apr) at (6.5,-0.62)  {$\icode{approve} \dblcolon \textbf{human}$};
  \node[Final] (e)   at (7.95,-0.62) {};
  \draw[E] (s) -- (tri) node[Lab,pos=0.5,above=9pt,inner sep=1pt]{(\kw{control})};
  \draw[E] (tri) -- (ans) node[Lab,pos=0.5,above=9pt,inner sep=1pt]{(\kw{cond}:\,faq)};
  \draw[E] (ans) -- (kb) node[Lab,pos=0.5,above=9pt,inner sep=1pt]{(\kw{tool})};
  \draw[E] (res) -- (rep) node[Lab,pos=0.5,below=9pt,inner sep=1pt]{(\kw{writes})};
  \draw[E] (rep) -- (apr) node[Lab,pos=0.5,below=9pt,inner sep=1pt]{(\kw{reads})};
  \draw[E] (apr) -- (e) node[Lab,pos=0.5,below=9pt,inner sep=1pt]{(\kw{control})};
  \draw[E] (tri) -- (res) node[Lab,midway,left,inner sep=3pt]{(\kw{cond}:\,complex)};
  \draw[E] (ans) -- (rep) node[Lab,midway,right,inner sep=3pt]{(\kw{writes})};
\end{tikzpicture}%
}
\end{center}
\end{example}

In \autoref{ex:workflow}, \icode{start} and \icode{end} mark the workflow boundaries, while \icode{triage} is a $\kw{router}$ step dispatching control along $\kw{cond}$ edges to the $\kw{agent}$ step \icode{answer} or the $\kw{subflow}$ step \icode{research}. The step \icode{answer} binds tool \icode{kb} via a $\kw{tool}$ edge, and both \icode{answer} and \icode{research} write to the shared channel \icode{reply}, whose reducer reconciles their updates. Workflows linked by $\kw{subflow}$ references form a \emph{program} $\mathcal{P} = \{W_1, \dots, W_k\}$ ($k = 1$ for a single graph), and $\mathsf{inBind}$ and $\mathsf{outBind}$ bind ports across their boundaries.

The behavioral labeling $\lambda$ distinguishes our IR from a dependence graph. Whereas a dependence graph records only that data flows into \icode{reply}, $\lambda$ and the channel reducers capture what each component means at runtime: whether an input is mandatory, how \icode{reply} merges concurrent writes, and which capabilities a role permits. As shown in \autoref{sec:taxonomy}, most workflow bugs are behavioral: the topology looks well-formed, and the fault lies in how the engine interprets a step or channel. Making $\lambda$ queryable lets one pattern language express both structural and behavioral bugs.

\subsection{The Predicate Vocabulary}
\label{sec:ir:predicates}

\tool encodes a workflow $W$ as ground relational facts over the \emph{base predicate} vocabulary $\Phi$ in \autoref{tab:predicates}. Where static lifting encounters dynamic constructs that cannot be resolved exactly, the front-end over-approximates the extracted facts so patterns report potential violations rather than silently missing them. Shared-state accesses ($\mathsf{writes}$ and $\mathsf{reads}$ on a channel, $\mathsf{store}$ and $\mathsf{inst}$ on a store or instance named by an opaque identity $\iota$) are modeled as predicates over state rather than edges in $E$.

\begin{table}[t]
\centering
\caption{The base predicate vocabulary $\Phi$.}
\label{tab:predicates}
\footnotesize
\renewcommand{\arraystretch}{1.18}
\setlength{\tabcolsep}{6pt}
\rowcolors{2}{white}{rowshade}
\begin{tabular}{ll}
\toprule
\textbf{Predicate} & \textbf{Meaning} \\
\midrule
$\mathsf{node}(v),\ \mathsf{kind}(v,\cdot)$ & step $v$ and its kind \\
$\mathsf{in}(s,v),\ \mathsf{out}(s,v)$ & input / output port $s$ of $v$ \\
$\mathsf{type}(s,\tau)$ & port $s$ has type $\tau$ \\
$\mathsf{default}(s),\ \mathsf{mandatory}(s)$ & port $s$ defaulted / required \\
$\mathsf{connOnly}(s)$ & port fillable only by a wired edge \\
$\mathsf{edge}(u,v,\delta),\ \mathsf{wired}(o,i)$ & edge of kind $\delta$; output $o$ wired to input $i$ \\
$\mathsf{branch}(u,v,\ell)$ & branch label on a \kw{cond} edge \\
$\mathsf{chan}(c),\ \mathsf{reducer}(c,\rho)$ & channel $c$ with reducer $\rho$ \\
$\mathsf{writes}(v,c),\ \mathsf{reads}(v,c)$ & $v$ writes / reads channel $c$ \\
$\mathsf{store}(v,\iota,\cdot)$ & $v$ reads/writes external store $\iota$ \\
$\mathsf{target}(v,t)$ & $v$ declares routing target $t$ \\
$\mathsf{subflow}(v,W')$ & $v$ embeds subflow $W'$ \\
$\mathsf{inBind}(v,\cdot),\ \mathsf{outBind}(v,\cdot)$ & cross-boundary port bindings \\
$\mathsf{name}(v,n),\ \mathsf{role}(v,\cdot)$ & name and role of $v$ \\
$\mathsf{start}(v),\ \mathsf{outputNode}(v)$ & entry / designated output node \\
$\mathsf{attr}(v,k,a)$ & attribute $k{=}a$ on $v$ ($\mathsf{attr}(v,k)$ flags $k$) \\
$\mathsf{execModel}(m)$ & $W$'s execution model is $m$ \\
$\mathsf{inst}(v,\iota)$ & in-memory object identity of $v$ \\
\bottomrule
\end{tabular}
\end{table}

The attribute predicate $\mathsf{attr}(v,k,a)$ provides an extensible key-value slot for the behavioral labels in $\lambda$. Front-ends map framework-specific flags onto generic semantic keys $k$, so engine-specific semantics enter the IR without expanding $\Phi$, and patterns never mention framework identifiers.

\subsection{Derived Relations}
\label{sec:ir:relations}

From $\Phi$, \tool computes the derived relations in
\autoref{fig:relations}. Factoring these recurring properties into reusable relations keeps the bug catalog declarative. The relations capture reachability ($\mathsf{reaches}$), strongly connected components ($\mathsf{scc}$), may-concurrency ($\mathsf{concurrent}$), and termination ($\mathsf{terminal}$), where a human checkpoint also counts as terminal because execution may pause there awaiting approval.

\begin{figure}[t]
\centering
{\footnotesize
\setlength{\jot}{2pt}
\begin{align*}
\mathsf{succ}(u,v) &\Leftarrow \mathsf{edge}(u,v,\_) \\
\mathsf{reaches}(u,v) &\Leftarrow \mathsf{succ}(u,v) \\
\mathsf{reaches}(u,v) &\Leftarrow \mathsf{succ}(u,w),\ \mathsf{reaches}(w,v) \\
\mathsf{scc}(C) &\triangleq \text{maximal } C \text{ s.t.\ } \forall\, u,v \in C.\\
&\quad\ u{=}v \vee (\mathsf{reaches}(u,v) \wedge \mathsf{reaches}(v,u)) \\
\mathsf{routingEdge}(v,t) &\Leftarrow \mathsf{edge}(v,t,\delta),\ \delta\in\{\kw{control},\kw{cond},\kw{handoff}\} \\
\mathsf{concurrent}(u,v) &\Leftarrow u\neq v,\ \neg\,\mathsf{reaches}(u,v),\ \neg\,\mathsf{reaches}(v,u) \\
\mathsf{concurrentW}(c) &\Leftarrow \mathsf{writes}(u,c),\ \mathsf{writes}(v,c),\ \mathsf{concurrent}(u,v) \\
\mathsf{terminal}(x) &\Leftarrow \mathsf{kind}(x,\kw{end}) \vee \mathsf{outputNode}(x) \vee \mathsf{kind}(x,\kw{human}) \\
\mathsf{assignable}(\tau,\tau') &\triangleq \text{execution-model-aware subtype check on $\tau \to \tau'$} \\
\mathsf{dist}(u,v) &\triangleq \text{shortest } \mathsf{succ}\text{-path length from } u \text{ to } v\text{, or } \infty \text{ if none}
\end{align*}
}
\caption{Relations derived over $\Phi$, some keyed to the execution model;
$\mathsf{dist}$ gives graph depth, used by the fan-in barrier check.}
\label{fig:relations}
\end{figure}

The subtype check $\mathsf{assignable}(\tau,\tau')$ depends on the execution model: under $\kw{dataflow}$, an output union is assignable to $\tau'$ when its members are contained in those of $\tau'$, whereas under $\kw{visual}$, an edge is coercive (the engine auto-converts the value). The types are
inherited from each framework: $\tau$ is the interface type the framework declares on a port, reused
as an opaque symbol and matched under that framework's own rules. The helper $\mathsf{fed}(s)$ holds when
an input port $s$ receives a value over a wired edge. Exactly
one node satisfies $\mathsf{start}(v)$; we write \textit{start} for it. Six auxiliary predicates capture local facts for specific categories: $\mathsf{resolves}(t)$ holds when target $t$ names a node anywhere in the program; $\mathsf{deliverable}(a,b)$ holds when a non-empty output of $a$ is $\mathsf{assignable}$ to an input of $b$; $\mathsf{bounded}(C)$ holds when cycle $C$ has a valid exit under the execution model; $\mathsf{andJoin}(j)$ holds when step $j$ fires only once all its predecessors arrive, and $\mathsf{unsat}(j)$ holds when they can never all arrive; and $\mathsf{declaresInput}(W')$ holds when subflow $W'$ requires an input port.

\section{Specifying \& Detecting Workflow Bugs}
\label{sec:bdl}

Once a workflow is lifted into ground facts over $\Phi$, detecting bugs reduces to querying those facts for violated safety properties. \tool expresses these properties in a declarative \emph{bug description language} (BDL) over $\Phi$ and the derived relations of \autoref{sec:ir:relations}, keeping the catalog open so that a new failure mode is a new rule rather than a new pass.

\subsection{The Bug Description Language}
\label{sec:bdl-lang}

A bug pattern is a Horn clause over the predicates $\Phi$ and the derived
relations:
\[
  \beta:\quad \mathsf{fire}(\bar x)\ \Leftarrow\ \phi_1,\dots,\phi_n.
\]
The body is built from atoms, negated atoms, and comparisons; an atom is
either a base predicate from the vocabulary $\Phi$ that \autoref{tab:predicates} lists, a
derived relation such as $\mathsf{reaches}$ or $\mathsf{concurrent}$, or the
theory predicate $\mathsf{assignable}$. The head $\mathsf{fire}(\bar x)$ binds
the \emph{witness} $\bar x$, the IR element (step, port, edge, target, channel, or cycle) held
responsible for the flaw. Because the witness is a variable of the body, a successful match reports a bug and pinpoints the exact IR element a developer must change. \autoref{fig:grammar} gives the grammar. As defined in \autoref{def:workflow},
$\mathsf{execModel}$ annotates a workflow with the scheduling and state-merging semantics of its
engine, and the analysis is parameterized by it: a clause carries it as a guard and is evaluated
under the semantics it is written for. The catalog is the finite set
$\mathcal B$ of patterns; each $\beta\in\mathcal B$ is a constrained Horn
clause~\cite{bjorner2015horn}, an implication that fires when its body holds,
evaluated over one program's finite IR.

\begin{figure}[t]
\footnotesize
\begin{align*}
  \beta &::= \mathsf{fire}(\bar x)\ \Leftarrow\ \mathit{body}\\
  \mathit{body}           &::= \mathit{lit}\ \mid\ \mathit{body}\,,\,\mathit{body}\ \mid\ \mathit{body}\wedge\mathit{body}\ \mid\ \mathit{body}\vee\mathit{body}\\
                          &\quad\mid\ (\mathit{body})\ \mid\ \exists\,x.\,\mathit{body}\ \mid\ \exists\,x\!\in\!S.\,\mathit{body}\\
  \mathit{lit}            &::= \phi\ \mid\ \neg\,\phi\ \mid\ \mathsf{execModel}(m)\ \mid\ t\mathbin{\circ}t\\
                          &\quad\ \mid\ S\mathbin{\circ_S}S\ \mid\ t\in S\ \mid\ t\notin S\ \mid\ \neg\,\exists\,x.\,\mathit{body}\\
  t                       &::= x\ \mid\ \mathit{const}\ \mid\ \varepsilon\ \mid\ |S|\ \mid\ \mathsf{dist}(t,t)\\
  S                       &::= X\ \mid\ \Sigma\ \mid\ \varnothing\ \mid\ \{\,x : \mathit{body}\,\}\\
  \circ                   &\in \{=,\neq,<,>,\leq,\geq\},\quad \circ_S \in \{=,\neq,\subseteq\}
\end{align*}
\noindent Here $\phi$ ranges over the base predicates of $\Phi$ in \autoref{tab:predicates}, the derived relations of \autoref{fig:relations}, and the auxiliary helpers of \autoref{sec:ir:relations}; comma and $\wedge$ denote conjunction, $X$ a set variable, $m$ an execution model, $\varepsilon$ an empty target, $\{\,x:\mathit{body}\,\}$ the set of $x$ satisfying $\mathit{body}$ with $|S|$ its size, and $\_$ a fresh existential variable.
\caption{The BDL grammar: a pattern is a Horn clause over the IR vocabulary, the
derived relations, and the theory predicate, with an optional execution-model gate.}
\label{fig:grammar}
\end{figure}

\subsection{The Bug-Pattern Catalog}
\label{sec:bdl-catalog}

\autoref{fig:catalog} gives the twelve detectors ($\beta_1$--$\beta_{12}$) of the catalog $\mathcal B$, grouped into
the six categories of \autoref{sec:taxonomy} and tagged structural, behavioral, or both. Where a
detector covers several failure modes, the figure shows its principal clauses. We walk through each
detector below alongside the minimal buggy IR fragment that triggers it.\footnote{\textbf{Visual notation}: Slate boxes denote steps, stone boxes denote shared state (channels, stores, or instances), ringed dots denote terminals, and dashed coral boxes denote nonexistent target nodes; port or channel types appear after \texttt{::}. Arrows are labeled with their corresponding IR predicates. Solid black arrows denote present edges, solid coral elements highlight present defects, and dashed coral elements indicate missing edges or unresolved references.}

\begin{figure*}[t]
\footnotesize
\setlength{\abovedisplayskip}{2pt}\setlength{\belowdisplayskip}{2pt}%
\setlength{\abovedisplayshortskip}{1pt}\setlength{\belowdisplayshortskip}{1pt}%
\setlength{\jot}{1.5pt}%
\hrule\vspace{2pt}
\noindent\textbf{Stalled \& Runaway Workflows}\hfill\textit{(structural \& behavioral)}\vspace{1pt}\hrule\vspace{1pt}
\begin{align*}
\blab{1}\fhead{s}   &\Leftarrow \mathsf{mandatory}(s),\ \mathsf{connOnly}(s),\ \neg\,\mathsf{fed}(s) & \textsc{starved-step}\\
\bnolab\fhead{j}    &\Leftarrow \mathsf{andJoin}(j),\ |\{p:\mathsf{routingEdge}(p,j)\}|\!\geq\!2,\ \mathsf{unsat}(j) & \textit{+ deadlocked join}\\
\blab{2}\fhead{C}   &\Leftarrow \mathsf{scc}(C),\ (|C|{>}1 \vee \exists\,v\!\in\!C.\ \mathsf{edge}(v,v,\_)),\ \neg\,\mathsf{bounded}(C) & \textsc{runaway-loop}\\
\intertext{\vspace{1pt}\hrule\vspace{1pt}\noindent\textbf{Broken Data Handoffs}\hfill\textit{(structural)}\vspace{1pt}\hrule\vspace{1pt}}
\blab{3}\fhead{u,v} &\Leftarrow \mathsf{out}(o,u),\ \mathsf{in}(i,v),\ \mathsf{wired}(o,i),\ \mathsf{type}(o,\tau),\ \mathsf{type}(i,\tau'),\ \neg\,\mathsf{assignable}(\tau,\tau') & \textsc{mistyped-handoff}\\
\bnolab\fhead{v}    &\Leftarrow \mathsf{execModel}(\kw{bsp}),\ \mathsf{writes}(v,c),\ \Sigma\neq\varnothing,\ c\notin\Sigma & \textit{+ undeclared channel write}\\
\bnolab\fhead{a,b}  &\Leftarrow \mathsf{execModel}(\kw{groupChat}),\ \mathsf{edge}(a,b,\_),\ \mathsf{out}(\_,a),\ \neg\,\mathsf{deliverable}(a,b) & \textit{+ message cannot route}\\
\intertext{\vspace{1pt}\hrule\vspace{1pt}\noindent\textbf{Misrouted Dispatch}\hfill\textit{(structural)}\vspace{1pt}\hrule\vspace{1pt}}
\blab{4}\fhead{v}   &\Leftarrow \mathsf{target}(v,t),\ t\neq\varepsilon,\ \neg\,\mathsf{resolves}(t) & \textsc{missing-route-target}\\
\bnolab\fhead{v,t}  &\Leftarrow \mathsf{execModel}(\kw{bsp}),\ \mathsf{edge}(v,t,\kw{control}),\ \big(\mathsf{target}(v,u) \vee \mathsf{edge}(v,u,\kw{cond})\big),\ u\neq t & \textit{+ double route}\\
\blab{5}\fhead{t}   &\Leftarrow \mathsf{branch}(r,t,\ell_1),\ \mathsf{branch}(r,t,\ell_2),\ \ell_1\neq\ell_2 & \textsc{ungated-branch-merge}\\
\blab{6}\fhead{t}   &\Leftarrow \mathsf{execModel}(\kw{visual}),\ \mathsf{branch}(r,t,\_),\ \neg\,\mathsf{settles}(t),\ \exists\,t'.\ \mathsf{branch}(r,t',\_)\wedge\mathsf{settles}(t') & \textsc{dead-end-branch}\\
\intertext{\vspace{1pt}\hrule\vspace{1pt}\noindent\textbf{Corrupted Shared State}\hfill\textit{(behavioral)}\vspace{1pt}\hrule\vspace{1pt}}
\blab{7}\fhead{c}   &\Leftarrow \mathsf{chan}(c),\ \mathsf{reducer}(c,\kw{lastValue}),\ \mathsf{concurrentW}(c) & \textsc{racing-channel-write}\\
\bnolab\fhead{v}    &\Leftarrow \mathsf{attr}(v,\kw{fullStateUpdate}),\ |\Sigma|\geq 2 & \textit{+ full-state overwrite}\\
\bnolab\fhead{d}    &\Leftarrow \mathsf{execModel}(\kw{bsp}),\ \neg\,\mathsf{attr}(d,\kw{barrier}),\ \mathsf{edge}(p_1,d,\kw{control}),\ \mathsf{edge}(p_2,d,\kw{control}), & \textit{+ fan-in barrier}\\
&\hphantom{{}\Leftarrow{}} \mathsf{concurrent}(p_1,p_2),\ \mathsf{dist}(\textit{start},p_1)\neq\mathsf{dist}(\textit{start},p_2) & \\
\blab{8}\fhead{a,b} &\Leftarrow \mathsf{inst}(a,\iota),\ \mathsf{inst}(b,\iota),\ a\neq b & \textsc{aliased-instance}\\
\blab{9}\fhead{r,w} &\Leftarrow \mathsf{store}(r,\iota,\kw{read}),\ \mathsf{store}(w,\iota,\kw{write}),\ \neg\,\mathsf{reaches}(w,r),\ r\neq w & \textsc{stale-read}\\
\intertext{\vspace{1pt}\hrule\vspace{1pt}\noindent\textbf{Unwired Subflows}\hfill\textit{(structural)}\vspace{1pt}\hrule\vspace{1pt}}
\blab{10}\fhead{v}  &\Leftarrow \mathsf{execModel}(\kw{visual}),\ \mathsf{subflow}(v,W'),\ \mathsf{edge}(v,\_,\kw{tool}),\ \mathsf{declaresInput}(W'),\ \neg\,\exists\,s.\ \mathsf{inBind}(v,s) & \textsc{unwired-subflow-input}\\
\intertext{\vspace{1pt}\hrule\vspace{1pt}\noindent\textbf{Role \& Interrupt Violations}\hfill\textit{(behavioral)}\vspace{1pt}\hrule\vspace{1pt}}
\blab{11}\fhead{v}  &\Leftarrow \big(\mathsf{role}(v,\kw{manager}) \vee (\mathsf{execModel}(\kw{groupChat})\wedge\mathsf{kind}(v,\kw{router}))\big),\ \mathsf{attr}(v,\kw{declaresTools}) & \textsc{over-privileged-coordinator}\\
\blab{12}\fhead{h_2} &\Leftarrow \mathsf{execModel}(\kw{visual}),\ \mathsf{kind}(h_1,\kw{human}),\ \mathsf{kind}(h_2,\kw{human}),\ \mathsf{reaches}(h_1,h_2), & \textsc{ungated-interrupt}\\
&\hphantom{{}\Leftarrow{}} \neg\,\mathsf{routerBetween}(h_1,h_2),\ h_1\neq h_2 & \\
\bnolab\fhead{h}    &\Leftarrow \mathsf{execModel}(\kw{visual}),\ \mathsf{kind}(h,\kw{human}),\ \mathsf{reaches}(\textit{start},h),\ \neg\,\exists\,a.\ \mathsf{kind}(a,\kw{agent})\wedge\mathsf{reaches}(a,h) & \textit{+ interrupt before agent}
\end{align*}
\hrule\vspace{3pt}
\noindent\textit{Helpers:}\ \ $\mathsf{settles}(t)\triangleq\mathsf{terminal}(t)\vee\exists y.\,\mathsf{reaches}(t,y)\wedge\mathsf{terminal}(y)$\qquad $\mathsf{routerBetween}(h_1,h_2)\triangleq\exists r.\,\mathsf{kind}(r,\kw{router})\wedge\mathsf{reaches}(h_1,r)\wedge\mathsf{reaches}(r,h_2)$
\caption{The \tool detector catalog $\mathcal B$ as Horn clauses over the IR, grouped
by category. Each entry is one \emph{detector}; its sub-clauses are the failure modes
that detector covers.}
\label{fig:catalog}
\end{figure*}

\subject{Stalled \& runaway workflows.} These two detectors check whether every step can fire and whether every loop can halt.
\begin{itemize}[topsep=3pt, leftmargin=1.5em]
\item \textsc{starved-step} ($\beta_1$).\bgraph{\begin{tikzpicture}
  \node[Pnode] (u) {$u$};
  \node[Pnode] (v) at (2.0,0) {$v$};
  \draw[M] (u) -- (v) node[LabC,above,midway]{(\kw{data})};
\end{tikzpicture}
}%
This pattern detects steps that cannot become ready or regions that cannot progress. Its principal clause flags a mandatory connection-only input port $s$ ($\mathsf{mandatory}(s)$, $\mathsf{connOnly}(s)$) fed by no edge ($\neg\mathsf{fed}(s)$), as shown above. Further clauses catch a step firing prematurely because all its inputs are defaulted, an AND-join ($\mathsf{andJoin}(j)$) whose predecessors can never all arrive ($\mathsf{unsat}(j)$), and a loop whose only entry is its own back-edge.
\item \textsc{runaway-loop} ($\beta_2$).\bgraph{\begin{tikzpicture}
  \node[Pnode] (a) {$a$};
  \node[Pnode] (b) at (1.6,0) {$b$};
  \node[Final] (t) at (3.1,0) {};
  \draw[E] (a) to[bend left=32] node[Lab,above,inner sep=0.6pt]{(\kw{control})} (b);
  \draw[E] (b) to[bend left=32] node[Lab,below,inner sep=0.6pt]{(\kw{control})} (a);
  \draw[M] (b) -- (t) node[LabC,above,midway]{(\kw{control})};
\end{tikzpicture}
}%
This pattern flags a nontrivial strongly connected component $C$ ($\mathsf{scc}(C)$) with no valid exit ($\neg\mathsf{bounded}(C)$): the inescapable loop that drained Devin's budget~\cite{husain2025devin}, as in the running example of \autoref{fig:ir}; the graph above is a cycle with no exit. Because termination is execution-model specific, a $\kw{visual}$ loop must be bounded by a structural loop node, whereas a $\kw{dataflow}$ cycle requires an escaping or may-emit edge.
\end{itemize}

\subject{Broken data handoffs.} This category adapts type checking to data edges and
shared channels.
\begin{itemize}[topsep=3pt, leftmargin=1.5em]
\item \textsc{mistyped-handoff} ($\beta_3$).\bgraph
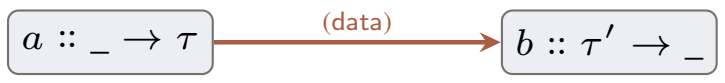
{\begin{tikzpicture}
  \node[Pnode] (a) {$a \dblcolon \_ \to \tau$};
  \node[Pnode] (b) at (3.3,0) {$b \dblcolon \tau' \to \_$};
  \draw[F] (a) -- (b) node[LabC,above,midway]{(\kw{data})};
\end{tikzpicture}
}%
In its principal clause, $\beta_3$ checks each wired port pair $\mathsf{wired}(o,i)$ and fires when the output type $\tau$ is not assignable to the input type $\tau'$ ($\neg\mathsf{assignable}(\tau,\tau')$), as shown above. Two further clauses cover shared state and messages: under $\kw{bsp}$, a step writing an undeclared channel ($c \notin \Sigma$), and under $\kw{groupChat}$, an agent sending a message to a peer with no matching handler ($\neg\mathsf{deliverable}(a,b)$).
\end{itemize}

\subject{Misrouted dispatch.} Because workflow routers compute their targets as data values rather than syntactic calls, routing defects manifest as unresolved targets and flawed branch topologies.
\begin{itemize}[topsep=3pt, leftmargin=1.5em]
\item \textsc{missing-route-target} ($\beta_4$).\bgraph
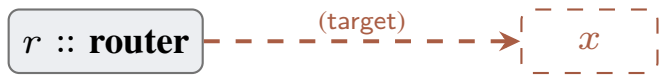
{\begin{tikzpicture}
  \node[Pnode] (r) {$r \dblcolon \textbf{router}$};
  \node[Ghost] (x) at (3.2,0) {$x$};
  \draw[M] (r) -- (x) node[LabC,above,midway]{(\kw{target})};
\end{tikzpicture}
}%
In its principal clause, $\beta_4$ checks each non-empty routing target $\mathsf{target}(v,t)$ against the whole program and fires if $t$ names no node ($\neg\mathsf{resolves}(t)$), the invalid-router-target defect of~\cite{ghAutogen781} shown above. Further clauses catch a router with no targets, a step routed two ways at once, and a stale route that reads an undeclared channel.
\item \textsc{ungated-branch-merge} ($\beta_5$).\bgraph
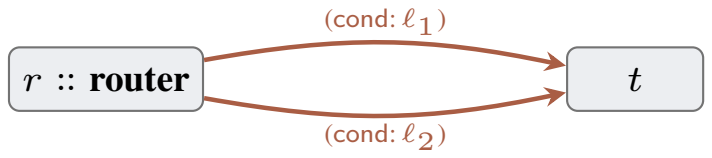
{\begin{tikzpicture}
  \node[Pnode] (r) {$r \dblcolon \textbf{router}$};
  \node[Pnode] (t) at (3.5,0) {$t$};
  \draw[F] (r) to[bend left=11] node[LabC,above,inner sep=1pt]{(\kw{cond}:\,$\ell_1$)} (t);
  \draw[F] (r) to[bend right=11] node[LabC,below,inner sep=1pt]{(\kw{cond}:\,$\ell_2$)} (t);
\end{tikzpicture}
}%
This detector flags a branch that runs even when not selected by its router; as shown above, the principal clause fires when two distinct branch labels $\ell_1 \neq \ell_2$ of a router $r$ converge on the same target $t$.
\item \textsc{dead-end-branch} ($\beta_6$).\bgraph
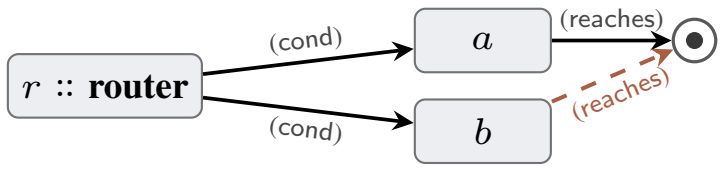
{\begin{tikzpicture}
  \node[Pnode] (r) {$r \dblcolon \textbf{router}$};
  \node[Pnode] (a) at (2.5,0.3) {$a$};
  \node[Pnode] (b) at (2.5,-0.3) {$b$};
  \node[Final] (e) at (3.9,0.3) {};
  \draw[E] (r) -- (a) node[Lab,sloped,above,pos=0.5,inner sep=1.3pt]{(\kw{cond})};
  \draw[E] (r) -- (b) node[Lab,sloped,below,pos=0.5,inner sep=1.3pt]{(\kw{cond})};
  \draw[E] (a) -- (e) node[Lab,above,midway,inner sep=1.3pt]{(\kw{reaches})};
  \draw[M] (b) -- (e) node[LabC,sloped,below,pos=0.5,inner sep=1.3pt]{(\kw{reaches})};
\end{tikzpicture}}%
This pattern detects a \emph{dangling branch}: a conditional branch $\mathsf{branch}(r,t,\_)$ whose target $t$ cannot reach any terminal ($\neg\mathsf{settles}(t)$) while a sibling branch $\mathsf{branch}(r,t',\_)$ does settle ($\mathsf{settles}(t')$), stalling any input routed to $t$.
\end{itemize}

\subject{Corrupted shared state.} Defects in this category are behavioral: the graph topology is well-formed, and the bug stems from concurrent or unordered state accesses.
\begin{itemize}[topsep=3pt, leftmargin=1.5em]
\item \textsc{racing-channel-write} ($\beta_7$).\bgraph
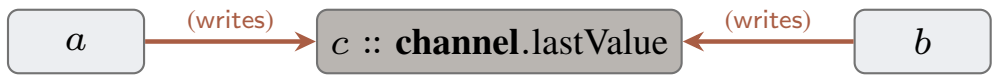
{\begin{tikzpicture}
  \node[Pnode] (a) {$a$};
  \node[Slot] (c) at (2.8,0) {$c \dblcolon \textbf{channel}.\text{lastValue}$};
  \node[Pnode] (b) at (5.6,0) {$b$};
  \draw[F] (a) -- (c) node[LabC,above,midway,inner sep=1.3pt]{(\kw{writes})};
  \draw[F] (b) -- (c) node[LabC,above,midway,inner sep=1.3pt]{(\kw{writes})};
\end{tikzpicture}
}%
This detector flags unsafe concurrent updates to shared state. Its principal clause matches a $\kw{lastValue}$ channel $c$ written by two may-concurrent steps ($\mathsf{concurrentW}(c)$), silently losing one update. Further clauses catch a full-state overwrite when multiple channels exist ($|\Sigma| \ge 2$) and a $\kw{bsp}$ fan-in without a barrier fed by concurrent branches at unequal depths.
\item \textsc{aliased-instance} ($\beta_8$).\bgraph
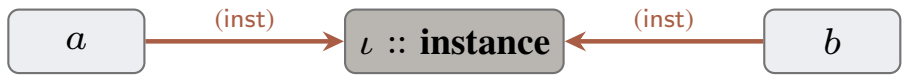
{\begin{tikzpicture}
  \node[Pnode] (a) {$a$};
  \node[Slot] (i) at (2.5,0) {$\iota \dblcolon \textbf{instance}$};
  \node[Pnode] (b) at (5.0,0) {$b$};
  \draw[F] (a) -- (i) node[LabC,above,midway,inner sep=1.3pt]{(\kw{inst})};
  \draw[F] (b) -- (i) node[LabC,above,midway,inner sep=1.3pt]{(\kw{inst})};
\end{tikzpicture}
}%
This pattern fires when two distinct steps $a \neq b$, within or across the graphs of $\mathcal{P}$, bind the same mutable instance $\iota$ ($\mathsf{inst}(a,\iota)$, $\mathsf{inst}(b,\iota)$), the workflow form of aliasing.
\item \textsc{stale-read} ($\beta_9$).\bgraph
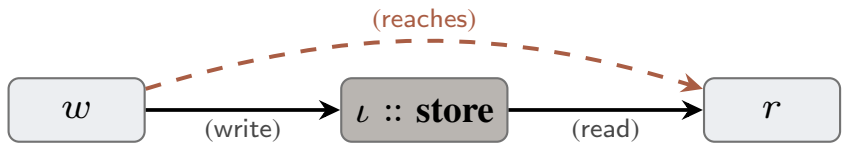
{\begin{tikzpicture}
  \node[Pnode] (w) {$w$};
  \node[Slot] (s) at (2.3,0) {$\iota \dblcolon \textbf{store}$};
  \node[Pnode] (r) at (4.6,0) {$r$};
  \draw[E] (w) -- (s) node[Lab,below,midway,inner sep=1.3pt]{(\kw{write})};
  \draw[E] (s) -- (r) node[Lab,below,midway,inner sep=1.3pt]{(\kw{read})};
  \draw[M] (w) to[bend left=17] node[LabC,above,midway,inner sep=1.5pt]{(\kw{reaches})} (r);
\end{tikzpicture}}%
This detector catches unordered read-after-write hazards on a shared store $\iota$: step $w$ writes $\iota$ and step $r$ reads $\iota$, with no path ordering the write before the read ($\neg\mathsf{reaches}(w,r)$).
\end{itemize}

\subject{Unwired subflows.} This category adapts inter-procedural binding to subflows that
embed one workflow inside another.
\begin{itemize}[topsep=3pt, leftmargin=1.5em]
\item \textsc{unwired-subflow-input} ($\beta_{10}$).\bgraph
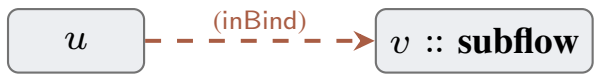
{\begin{tikzpicture}
  \node[Pnode] (u) {$u$};
  \node[Pnode] (sf) at (2.7,0) {$v \dblcolon \textbf{subflow}$};
  \draw[M] (u) -- (sf) node[LabC,above,midway]{(\kw{inBind})};
\end{tikzpicture}
}%
This inter-workflow pattern fires when a step $v$ embeds a subflow $W'$ as a tool ($\mathsf{subflow}(v,W')$, $\mathsf{edge}(v,\_,\kw{tool})$) that requires an input port ($\mathsf{declaresInput}(W')$), yet $v$ leaves it unbound ($\neg\exists s.\,\mathsf{inBind}(v,s)$).
\end{itemize}

\subject{Role \& interrupt violations.} These detectors query the behavioral labeling
$\lambda$ to enforce role and protocol contracts.
\begin{itemize}[topsep=3pt, leftmargin=1.5em]
\item \textsc{over-privileged-coordinator} ($\beta_{11}$).\bgraph
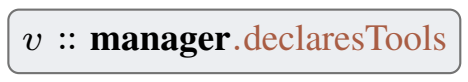
{\begin{tikzpicture}
  \node[Pnode] (v) {$v \dblcolon \textbf{manager}\textcolor{archCoral!78!black}{.\text{declaresTools}}$};
\end{tikzpicture}
}%
This detector flags capability over-reach: a step $v$ with a coordinating role ($\mathsf{role}(v,\kw{manager})$ or a $\kw{groupChat}$ router) also declares tools ($\mathsf{attr}(v,\kw{declaresTools})$), violating its role contract.
\item \textsc{ungated-interrupt} ($\beta_{12}$).\bgraph
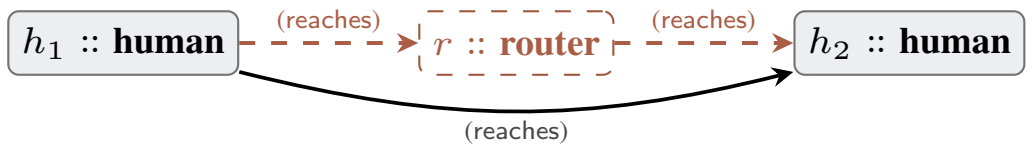
{\begin{tikzpicture}
  \node[Pnode] (h1) {$h_1 \dblcolon \textbf{human}$};
  \node[Ghost] (rt) at (2.6,0) {$r \dblcolon \textbf{router}$};
  \node[Pnode] (h2) at (5.2,0) {$h_2 \dblcolon \textbf{human}$};
  \draw[M] (h1) -- (rt) node[LabC,above,midway,inner sep=1.5pt]{(\kw{reaches})};
  \draw[M] (rt) -- (h2) node[LabC,above,midway,inner sep=1.5pt]{(\kw{reaches})};
  \draw[E] (h1) to[bend right=14] node[Lab,below,midway,inner sep=1.5pt]{(\kw{reaches})} (h2);
\end{tikzpicture}}%
This pattern governs human-in-the-loop ordering: its principal clause flags two human checkpoints $h_1, h_2$ in sequence with no router between ($\neg\mathsf{routerBetween}(h_1,h_2)$), serializing approvals meant to be conditional~\cite{ghLanggraphjs792,ghDeepagents554}; a second clause flags an interrupt reachable from $\textit{start}$ with no preceding agent.
\end{itemize}

The behavioral detectors, those whose bodies query $\lambda$ rather than topology and
types alone, cover the failure modes a purely structural view cannot express.

\section{Implementation}
\label{sec:impl}

\tool is an open-source, framework-neutral analysis backend, about $4{,}900$ lines of Rust, with one front-end per supported framework. All reasoning lives in the backend, which builds the IR, computes derived relations, and matches the catalog $\mathcal{B}$ as Horn clauses over one program's facts. The clauses can be checked by an off-the-shelf Datalog engine such as Souffl\'e~\cite{jordan2016souffle}, but calling an external solver is too slow, so we built a customized checker in the Rust backend as an optimization.

Each front-end reads the workflow object the framework has built. Based on each framework's documentation and on how frequently developers use each API, we support over $150$ builder APIs and node types across the six frameworks. As in classical static program analysis, where all potential edges and targets can be resolved at build time, for instance in dynamic routing and fan-out, we resolve every such edge and target and treat the choice among those targets as non-deterministic. Likewise, where a parent task spawns a sub-workflow that is also available at build time, we construct the parent and the sub-workflow together, so the case reduces to a single workflow. Where an edge cannot be resolved at build time, we conservatively assume it is unsafe.

\section{Evaluation}
\label{sec:eval}

Three research questions guide our evaluation:

\begin{itemize}[noitemsep, topsep=2pt, leftmargin=1.5em]
\item \textbf{RQ1 (Effectiveness and Cost).} How accurately and efficiently does
      \tool detect workflow bugs across frameworks, and how does it compare against
      existing baselines?
\item \textbf{RQ2 (Ablation).} How much does behavioral reasoning contribute to
      bug detection beyond structural analysis?
\item \textbf{RQ3 (Previously Unknown Bugs).} Can \tool uncover previously unknown
      workflow bugs in deployed projects?
\end{itemize}

\subsection{Experimental Setup}
\label{sec:eval:protocol}

\subject{Benchmarks and Selection Criteria.} From our corpus, we constructed a suite of $135$ reproducible, labeled benchmarks across the six frameworks supported by our front-ends (\autoref{tab:main}). Inspecting each framework's issue tracker and pull requests, we retained a report only if (1) its root cause lies in the workflow's wired logic and leaves a static graph signature, (2) it is reproducible from a minimal workflow artifact, and (3) it has a maintainer-accepted fix. We excluded reports caused by environment configuration or model quality, as well as those lacking a reproducible artifact or fix. Each benchmark thus pairs a \emph{buggy} workflow artifact with its \emph{fixed} counterpart, so an ideal analyzer should report the defect on the buggy version and produce no warning on the fixed version. Ground-truth bug status is determined by the issue reports, while category and pattern labels were assigned independently by three graduate students and retained upon majority agreement.

\subject{Metrics.} We evaluate bug detection as binary classification, treating each \emph{buggy}
artifact as a positive instance (\#Bug in \autoref{tab:main}) and its \emph{fixed} counterpart as a negative instance (\#Fix). For each framework and overall, we report precision $P$, recall $R$, harmonic mean $F_1$, and false-positive rate $\FPR$.

\subsection{RQ1: Effectiveness and Cost}
\label{sec:eval:rq1}

\begin{table}[t]
  \centering
  \caption{Per-framework and overall detection of \tool.}
  \label{tab:main}
  \footnotesize
  \setlength{\tabcolsep}{3pt}
  \rowcolors{2}{white}{rowshade}
  \begin{tabular}{l rrr | cccc}
    \toprule
    Framework & \#Bug & \#Fix & \#Tot & $P$ & $R$ & $F_1$ & \FPR \\
    \midrule
    Haystack~\cite{haystack}              & 25 & 25 & 50  & 0.92 & 0.88 & 0.90 & 0.08 \\
    LangGraph~\cite{langgraph}            & 33 & 33 & 66  & 0.96 & 0.76 & 0.85 & 0.03 \\
    Flowise~\cite{flowise}                & 24 & 24 & 48  & 0.83 & 0.83 & 0.83 & 0.17 \\
    Langflow~\cite{langflow}              & 13 & 13 & 26  & 0.86 & 0.92 & 0.89 & 0.15 \\
    AutoGen~\cite{wu2023autogen}          & 14 & 14 & 28  & 0.92 & 0.79 & 0.85 & 0.07 \\
    Agent Framework~\cite{agentframework} & 26 & 26 & 52  & 0.96 & 0.88 & 0.92 & 0.04 \\
    \midrule
    Overall           & 135 & 135 & 270 & 0.91 & 0.84 & 0.87 & 0.08 \\
    \bottomrule
  \end{tabular}
\end{table}

\autoref{tab:main} summarizes \tool's detection accuracy across the six frameworks. Overall, \tool achieves $0.91$
precision, $0.84$ recall, and $F_1{=}0.87$, with $\FPR{=}0.08$. Across frameworks, recall is highest on
Langflow ($R{=}0.92$), whose exported flows expose readiness and type contracts, while
LangGraph and Agent Framework reach the highest precision ($P{=}0.96$, $\FPR{=}0.03$/$0.04$). Recall is lower on LangGraph ($0.76$) and AutoGen ($0.79$).

\subject{Comparison to baselines.} \autoref{tab:baselines} compares \tool against three classes of baselines on the same benchmarks. Where a baseline does not natively support a target framework, we extend it to ensure a fair comparison.

First, a \emph{framework-native validator} runs the built-in type checker and graph compiler shipped by each framework. All $135$ buggy
workflows pass native validation ($R{=}0.00$), as these defects are well-typed at the host level and native validators do not check single-branch routes, concurrent state merges, or role capabilities, while serving only one framework.

Second, to evaluate a \emph{general static analyzer}, we run
CodeQL~\cite{codeql} over the host Python code. Because a workflow graph is constructed dynamically through builder API calls rather than represented as a first-class syntactic construct, standard syntactic queries recover very few bugs ($R{\approx}0.07$) and do not apply to JSON-exported visual flows. Even when we build a custom structural analyzer
(\emph{CodeQL Extended}) that parses exported graph structures directly, it detects only purely topological flaws ($R{=}0.17$), missing the three-quarters of defects that depend on step-level execution semantics.

Third, we evaluate three \emph{LLM-as-judge} baselines (GPT-5.5~\cite{gpt55}, Opus-4.8~\cite{opus48},
and DeepSeek-V4~\cite{deepseekv4}) prompted with our six-category taxonomy; \appref{app:llm-prompt} gives the prompt. While these models achieve high recall ($R{=}0.77$--$0.88$), their false-positive rates reach $0.49$--$0.67$. Requiring the correct defect category (\emph{Strict} rows) roughly halves their recall (\mbox{$0.39$--$0.44$}).

\begin{table}[t]
  \centering
  \caption{Baseline comparison over the benchmark suite, \tool also broken down by bug category.
  \emph{Strict} rows require the reported category to match the ground truth; the three LLM
  judges run at \textsc{x-high} reasoning or its equivalent.}
  \label{tab:baselines}
  \footnotesize
  \rowcolors{2}{white}{rowshade}
  \begin{tabular}{l cccc}
    \toprule
    Approach & $P$ & $R$ & $F_1$ & \FPR \\
    \midrule
    Framework-Native Validator & --- & 0.00 & --- & 0.00 \\
    CodeQL~\cite{codeql}          & 0.71 & 0.07 & 0.13 & 0.03 \\
    CodeQL~\cite{codeql} Extended & 0.74 & 0.17 & 0.28 & 0.06 \\
    LLM (GPT-5.5~\cite{gpt55})          & 0.60 & 0.80 & 0.68 & 0.54 \\
    LLM (GPT-5.5~\cite{gpt55}) Strict    & 0.44 & 0.43 & 0.44 & 0.54 \\
    LLM (Opus-4.8~\cite{opus48})        & 0.61 & 0.77 & 0.68 & 0.49 \\
    LLM (Opus-4.8~\cite{opus48}) Strict  & 0.45 & 0.39 & 0.42 & 0.49 \\
    LLM (DeepSeek-V4~\cite{deepseekv4}) & 0.57 & 0.88 & 0.69 & 0.67 \\
    LLM (DeepSeek-V4~\cite{deepseekv4}) Strict & 0.40 & 0.44 & 0.42 & 0.67 \\
    \midrule
    \textbf{\tool\ (Our Tool)}    & \textbf{0.91} & \textbf{0.84} & \textbf{0.87} & \textbf{0.08} \\
    \quad Stalled \& Runaway Workflows & 0.94 & 0.81 & 0.87 & 0.05 \\
    \quad Broken Data Handoffs  & 1.00 & 0.84 & 0.91 & 0.00 \\
    \quad Misrouted Dispatch   & 0.81 & 0.88 & 0.85 & 0.20 \\
    \quad Corrupted Shared State    & 0.93 & 0.81 & 0.87 & 0.06 \\
    \quad Unwired Subflows             & 0.92 & 0.85 & 0.88 & 0.08 \\
    \quad Role \& Interrupt Violations  & 0.89 & 0.89 & 0.89 & 0.11 \\
    \bottomrule
  \end{tabular}
\end{table}

Overall, \tool outperforms the strongest baseline by $0.18$ in $F_1$. Whereas baselines are confined to one framework, limited to surface topology, or prone to high false-positive rates, \tool achieves high precision and recall by matching declarative patterns over a behaviorally annotated IR.
\autoref{tab:baselines} also breaks \tool's accuracy down by category: precision of $0.81$--$1.00$, recall of $0.81$--$0.89$, and $\FPR \le 0.20$.

\subject{Analysis cost.} \tool analyzes a workflow end-to-end in a median of $0.7$\,s (mean $0.8$\,s), dominated by
front-end IR lifting and well within the budget of a pre-deployment CI check.

\takeaways{\tool detects workflow bugs with $0.91$ precision and $0.84$ recall ($F_1{=}0.87$) in $0.7$\,s per workflow, outperforming the strongest baseline by $0.18$ $F_1$ and pinpointing each defect.}

\begin{table*}[t]
\centering
\caption{Selected previously unknown workflow bugs found by \tool, all fixed by their maintainers; \appref{app:wild} lists them all.}
\label{tab:wild}
\footnotesize
\renewcommand{\arraystretch}{0.95}
\setlength{\tabcolsep}{2pt}
\rowcolors{2}{white}{rowshade}
\begin{tabular}{r p{7.5cm} l l l l}
\toprule
\# & Bug description & Category & Framework & Issue & Status \\
\midrule
1 & Crew is built from component state reused across requests & Corrupted Shared State & Langflow & \href{https://github.com/langflow-ai/langflow/issues/13846}{langflow \#13846} & Fixed \\
2 & One agent wired into several tasks is mutated in place & Corrupted Shared State & Langflow & \href{https://github.com/langflow-ai/langflow/issues/13847}{langflow \#13847} & Fixed \\
3 & Agent factory passes an unconverted model to the crew agent & Broken Data Handoffs & Langflow & \href{https://github.com/langflow-ai/langflow/issues/13848}{langflow \#13848} & Fixed \\
4 & Agent router returns a label absent from the path map & Misrouted Dispatch & LangChain & \href{https://github.com/langchain-ai/langchain/issues/38351}{langchain \#38351} & Fixed \\
5 & Shared router returns targets absent from the edge's path map & Misrouted Dispatch & LangGraph & \href{https://github.com/TauricResearch/TradingAgents/issues/1088}{TradingAgents \#1088} & Fixed \\
6 & Resume reuses a checkpoint from a differently shaped graph & Corrupted Shared State & LangGraph & \href{https://github.com/TauricResearch/TradingAgents/issues/1089}{TradingAgents \#1089} & Fixed \\
7 & Run config drops the thread id on the new context path & Broken Data Handoffs & LangGraph & \href{https://github.com/bytedance/deer-flow/issues/3728}{deer-flow \#3728} & Fixed \\
8 & Tools bind the memory bank at build time, so tenants share it & Corrupted Shared State & LangGraph & \href{https://github.com/vectorize-io/hindsight/issues/2442}{hindsight \#2442} & Fixed \\
9 & Fan-in wired as two edges, not a barrier; fires twice & Corrupted Shared State & LangGraph & \href{https://github.com/MODSetter/SurfSense/issues/1531}{SurfSense \#1531} & Fixed \\
10 & Sub-agent failure counter resets on rewrap; cap unreachable & Stalled \& Runaway Workflows & LangGraph & \href{https://github.com/PurpleAILAB/Decepticon/issues/722}{Decepticon \#722} & Fixed \\
11 & Parallel writes to a channel with no reducer (lost update) & Corrupted Shared State & LangGraph & \href{https://github.com/NVIDIA/NeMo-Agent-Toolkit/issues/2084}{NVIDIA / NeMo \#2084} & Fixed \\
12 & Unconditional edge and router both fire; \icode{END} unreachable & Misrouted Dispatch & LangGraph & \href{https://github.com/NVIDIA/NeMo-Agent-Toolkit/issues/2083}{NVIDIA / NeMo \#2083} & Fixed \\
13 & A final node has no edge to \icode{END}; never terminates & Stalled \& Runaway Workflows & LangGraph & \href{https://github.com/NVIDIA/NeMo-Agent-Toolkit/issues/2082}{NVIDIA / NeMo \#2082} & Fixed \\
14 & Loop's step-count guard is unreachable; iterations unbounded & Stalled \& Runaway Workflows & LangGraph & \href{https://github.com/starpig1129/DATAGEN/issues/29}{DATAGEN \#29} & Fixed \\
15 & Final refinement is returned without passing the reviewer & Role \& Interrupt Violations & Agent Framework & \href{https://github.com/microsoft/OpenAIWorkshop/issues/432}{OpenAIWorkshop \#432} & Fixed \\
\bottomrule
\end{tabular}
\end{table*}

\subsection{RQ2: Ablation}
\label{sec:eval:rq2}

To quantify the contribution of the IR's behavioral annotations, we compare the full \tool catalog (\emph{structural$+$behavioral}) against an ablated variant restricted to \emph{structural-only} patterns. \autoref{fig:ablation} reports per-category recall for both variants, with each bug grouped by its ground-truth category and credited to whichever detector fires.

\begin{figure}[t]
\centering
\begin{tikzpicture}
\pgfplotstableread[col sep=space]{
    label Struct Behav
    Stall 0.24 0.81
    Handoff 0.79 0.84
    Route 0.64 0.88
    State 0.31 0.81
    Subflow 0.77 0.85
    Role 0.78 0.89
}\abldata
\begin{axis}[
    ybar, bar width=11pt, bar shift auto=false,
    area legend,
    width=\columnwidth, height=3.6cm,
    axis lines=box,
    axis line style={draw=black!45, line width=0.5pt},
    ymin=0, ymax=1.36,
    ytick={0,0.2,0.4,0.6,0.8,1.0},
    yticklabel style={font=\sffamily\scriptsize\bfseries},
    symbolic x coords={Stall,Handoff,Route,State,Subflow,Role},
    xtick=data,
    xticklabel style={font=\sffamily\scriptsize\bfseries, yshift=2pt},
    enlarge x limits=0.09,
    tick style={draw=black!45},
    tick align=outside,
    major tick length=2pt,
    ymajorgrids, grid style={dashed, gray!28, line width=0.4pt},
    legend style={at={(0.5,-0.20)},anchor=north,legend columns=2,draw=none,
        font=\sffamily\scriptsize\bfseries,column sep=1.2ex},
    legend cell align=left,
    legend image code/.code={\fill (0cm,-0.05cm) rectangle (0.34cm,0.11cm);},
    nodes near coords={\textpdfrender{TextRenderingMode=2,LineWidth=0.2pt}{\pgfmathprintnumber[fixed,fixed zerofill,precision=2]{\pgfplotspointmeta}}},
    every node near coord/.append style={font=\sffamily\fontsize{5pt}{5pt}\selectfont\bfseries, rotate=90, anchor=west,
        inner sep=1.0pt, text=black},
    clip=false,
]
\addplot[fill=archSlate!30, draw=none, bar shift=-5.5pt] table[x=label, y=Struct] {\abldata};
\addplot[fill=archSlate, draw=none, bar shift=5.5pt] table[x=label, y=Behav] {\abldata};
\node[anchor=south east, font=\sffamily\scriptsize\bfseries, text=black!72, inner sep=0pt]
    at ([xshift=-4pt,yshift=2pt]rel axis cs:0,0.80) {recall};
\legend{Structural only, {+ Behavioral}}
\end{axis}
\end{tikzpicture}
\caption{Per-category recall under the structural-only catalog versus the full
structural$+$behavioral catalog; the behavioral layer recovers the recall topology alone misses.}
\label{fig:ablation}
\end{figure}

Incorporating behavioral patterns detects $46$ additional bugs missed by topology alone, raising overall
recall from $0.50$ to $0.84$. The largest gains occur where defects hinge on step-level execution semantics: recall jumps from $0.24$ to $0.81$ in Stalled \& Runaway Workflows and from $0.31$ to $0.81$ in Corrupted Shared State once input readiness flags (\icode{mandatory}, \icode{default}) and channel reducers become visible, while Misrouted Dispatch rises from $0.64$ to $0.88$. In the remaining categories, structural patterns capture most wiring flaws, and behavioral predicates raise recall on Broken Data Handoffs ($0.79{\to}0.84$), Unwired Subflows ($0.77{\to}0.85$), and Role \& Interrupt Violations ($0.78{\to}0.89$).

\takeaways{Structural patterns alone achieve $0.50$ recall; adding the IR's behavioral annotations detects $46$ additional bugs and raises overall recall to $0.84$.}

\subsection{RQ3: Previously Unknown Bugs}
\label{sec:eval:rq3}

We adapted \tool to deployed, non-corpus workflows across more than $40$ actively maintained open-source agent projects, including NVIDIA's NeMo Agent Toolkit~\cite{nemoagent},
LangChain~\cite{langchain}, and CopilotKit~\cite{copilotkit}. \tool uncovered \textbf{36} previously unknown workflow bugs across $21$ repositories carrying hundreds of thousands of GitHub stars, all confirmed or fixed by their
maintainers ($33$ fixed, $3$ confirmed). \autoref{tab:wild} lists a selection; \appref{app:wild} lists them all. Several sit in the framework's own examples that developers copy into production and pass the framework's own checks.

\subject{Case A: Lost update from a missing channel reducer.} NVIDIA's NeMo Agent Toolkit ships a map-reduce example that fans a batch out to parallel \icode{create\_batch\_summary} steps,
each writing to a shared \icode{summaries} channel declared as a plain list without a
reducer. Under the engine's bulk-synchronous schedule, concurrent branches finishing in the same
superstep collide on \icode{summaries}: one write silently overwrites the other, or the runtime
aborts with \icode{InvalidUpdateError}. Because the workflow is well-typed, it compiles cleanly
and fails only on batches large enough to trigger concurrent writes. \tool's $\beta_{7}$ statically flags the concurrent writes to
\icode{summaries}, prompting the maintainers to add a
reducer.\footnote{\url{https://github.com/NVIDIA/NeMo-Agent-Toolkit/issues/2084}}

\subject{Case B: Unresolved route in a conditional path map.} In LangChain's \icode{create\_agent},
the model step is followed by a conditional edge whose path map registers the tool step and the exit
node, but includes the model step itself only under certain configurations. However, the router function still
returns the model step whenever all pending tool calls have already been resolved in a single turn
(for example, when a middleware injects cached or pre-approved tool outputs). In the default
configuration, this return label is absent from the path map, causing the runtime to crash with
\icode{KeyError(\textquotesingle model\textquotesingle)}. \tool compares the router's declared return targets against the path map and flags the unresolved \icode{model} target; the maintainers merged a fix adding the missing
route.\footnote{\url{https://github.com/langchain-ai/langchain/issues/38351}}

\takeaways{On deployed real-world workflows, \tool uncovered \textbf{36} previously unknown bugs across $21$ repositories ($33$ fixed and $3$ confirmed by maintainers).}

\subject{Threats to validity.} We observe two threats in particular. The frameworks evolve quickly
and their engine semantics drift across releases; so far the drift has stayed within what the IR
expresses, so a front-end absorbs a release while the IR and detectors stand, and the suite
pins each benchmark to its version. New frameworks also keep appearing, which the shared IR meets with
one front-end each.

\section{Related Work}
\label{sec:related}

\begin{table*}[t]
\centering
\caption{How agentic workflows typically differ from business-process workflows, and the
detectors that account for each difference.}
\label{tab:bpm}
\footnotesize
\renewcommand{\arraystretch}{0.95}
\rowcolors{2}{white}{rowshade}
\begin{tabular}{p{2.2cm} >{\raggedright\arraybackslash}p{5.2cm} >{\raggedright\arraybackslash}p{7.4cm} l}
\toprule
\textbf{Aspect} & \textbf{Business process} & \textbf{Agentic workflow} & \textbf{Detectors} \\
\midrule
Control transfer &
Transition targets are fixed in the process definition. &
The next step can be a name the model returns at run time, which may match no step. &
$\beta_4$--$\beta_6$ \\
Termination &
A loop exits on a condition the designer wrote over case data. &
A loop may exit only on a value the model produces, so a bound has to be added explicitly. &
$\beta_2$ \\
Authority &
A person or program performs a task under an assigned role. &
The model chooses its own tool calls, so which tools a step holds and where approvals sit
become properties of the graph. &
$\beta_{11}$, $\beta_{12}$ \\
Shared state &
A data element holds one value, so concurrent writes conflict. &
Each channel has a merge rule (a reducer), so the same concurrent writes either merge or
lose an update. &
$\beta_7$ \\
\bottomrule
\end{tabular}
\end{table*}

\tool builds on four lines of related research.

\subject{Failure studies of LLM agents.}
Several empirical studies characterize how deployed agent systems fail in practice. Taxonomies such as
MAST~\cite{cemri2025mast} categorize failed multi-agent execution traces into recurring modes of
mis-coordination, dropped steps, and premature termination, while incident reports document
severe real-world losses from runaway loops and unguarded actions~\cite{husain2025devin, claburn2025devin, zenity2026pocketos,
gupta2025gemini}. Across these studies, the critical failures stem from how workflow steps are wired and coordinated more than from any single model invocation.
However, empirical studies and post-mortem analyses diagnose failures only \emph{after} execution, once damage has occurred.
In contrast, \tool detects these failures statically before deployment by formalizing recurring failure modes as declarative patterns over a workflow's IR.

\subject{Safeguards for agent frameworks.}
Existing safeguards for agentic workflows fall into two categories.
\emph{Framework-native checks} validate basic graph construction: for example, Haystack rejects
connections between ports with incompatible types~\cite{haystack}, while
LangGraph and Microsoft's Agent Framework provide typed state schemas and routing
primitives~\cite{langgraph, agentframework}. \emph{Dynamic guardrails} operate at
runtime: programmable rails constrain agent outputs~\cite{rebedea2023nemo},
and LLM-as-judge evaluators score execution traces probabilistically~\cite{zheng2023judging}.
However, native validators miss behavioral defects and apply only to a single framework, whereas
dynamic guardrails intervene only during or after execution. \tool bridges this gap by performing deterministic,
cross-framework static analysis over a behaviorally annotated IR prior to deployment.

\subject{Verification of business-process workflows.}
Workflow verification has been studied extensively in business process management and
service-oriented computing. Workflow nets model a process as a Petri
net~\cite{murata1989petri}, and their \emph{soundness} property requires that every case can
complete, completes properly, and leaves no task
dead~\cite{vanderaalst1998workflow, vanderaalst2011soundness}.
Related lines of work catalog the control-flow constructs of process languages and the ways a
process handles exceptions~\cite{vanderaalst2003patterns, russell2006exception}, detect
data-flow anti-patterns such as reading data that is not yet
available~\cite{trcka2009dataflow}, enforce authorization constraints on who may perform a
task~\cite{bertino1999authorization}, and analyze processes written in BPEL and
BPMN~\cite{fu2004bpel, dijkman2008bpmn}. Our bug categories share conceptual roots with these classical properties. When a router's decision is treated as a free
choice~\cite{desel1995freechoice}, stalled and runaway workflows and misrouted dispatch relate
to soundness, the three data-centric categories (broken data handoffs, corrupted shared state,
and unwired subflows) to data-flow anti-patterns, and role and interrupt violations to
authorization and exception patterns. The agentic setting differs in where the decisions come
from. In a business process, the designer typically fixes transition targets, exit conditions,
and task assignments in the process definition, whereas an agentic workflow leaves some of
these to a language model at run time and merges shared state through per-channel reducers.
\autoref{tab:bpm} summarizes these differences together with the detectors that account for
them.

\subject{Static analysis and bug-pattern languages.}
General-purpose static analyzers and query engines detect defects by matching
declarative rules against a program abstraction; CodeQL over a code property
graph~\cite{codeql}, Infer~\cite{infer}, Semgrep~\cite{semgrep}, and Datalog-based
whole-program analyses~\cite{bravenboer2009doop} are representative. \tool adopts this methodology, a graph abstraction paired with a pattern language, and extends it in two
ways that agentic workflows require. First, because a workflow is built through heterogeneous host-language
APIs and visual builders rather than a single program text, recovering it as a semantic IR is part of
the problem. Second, because its defining failures hinge on step-level behavior rather than topology
alone, the pattern language quantifies over behavioral labels.

\section{Conclusion}
\label{sec:conclusion}

We presented \tool, a static safety analyzer that detects \emph{workflow bugs} in agentic workflows before deployment. Grounded in a study of $861$ issue reports across $13$ orchestration frameworks, we derived a six-category defect taxonomy, constructed the first labeled corpus of $783$ workflow bugs, and assembled a reproducible suite of $135$ benchmarks. To analyze workflows across frameworks, \tool lifts framework-specific definitions into a unified intermediate representation capturing both graph topology and step-level behavioral semantics, and checks declarative bug patterns expressed in a Horn-clause bug description language. On our benchmarks, \tool achieves $91\%$ precision and $84\%$ recall ($F_1{=}0.87$) in $0.7$\,s per workflow, outperforming the strongest baseline by $0.18$~$F_1$, with behavioral reasoning recovering $46$ bugs missed by structural checks alone. In deployed open-source projects, \tool uncovered \textbf{36} previously unknown workflow bugs, $33$ of which have already been fixed by maintainers.

\section*{Data Availability}
\tool and the benchmark suite are available at
\url{https://doi.org/10.5281/zenodo.24408668}.

\section*{Acknowledgment}
We are truly grateful for the time and effort that the anonymous reviewers invested in reviewing our
work and offering valuable feedback. This work is supported in part by a Google Faculty Research
Award and an Ethereum Foundation Academic Award. For authors, Ying Li is supported by Google PhD
Fellowship. The UCLA authors are supported in part by NSF grants 2531140, 2317184, and 2411153,
and by \mbox{Amazon}, Cisco, Coefficient Giving, and Keysight. Any opinions, findings, and conclusions or recommendations expressed in this work are
those of the author(s) and do not necessarily reflect the views of the institutions above. We used
Claude (Anthropic) throughout this paper to assist with writing and editing.

\IEEEtriggeratref{55}
\bibliographystyle{IEEEtran}
\bibliography{references}

\clearpage
\twocolumn[{{\centering
\expandafter\def\csname @captype\endcsname{table}%
\caption{All previously unknown workflow bugs \tool surfaced in deployed open-source projects that
have been fixed or confirmed by their maintainers.}
\label{tab:wild-full}
\footnotesize
\renewcommand{\arraystretch}{0.95}
\setlength{\tabcolsep}{2pt}
\rowcolors{2}{white}{rowshade}
\begin{tabular}{r p{7.2cm} l l l l}
\toprule
\# & Bug description & Category & Framework & Issue & Status \\
\midrule
1 & Crew is built from component state reused across requests & Corrupted Shared State & Langflow & \href{https://github.com/langflow-ai/langflow/issues/13846}{langflow \#13846} & Fixed \\
2 & One agent wired into several tasks is mutated in place & Corrupted Shared State & Langflow & \href{https://github.com/langflow-ai/langflow/issues/13847}{langflow \#13847} & Fixed \\
3 & Agent factory passes an unconverted model to the crew agent & Broken Data Handoffs & Langflow & \href{https://github.com/langflow-ai/langflow/issues/13848}{langflow \#13848} & Fixed \\
4 & Agent router returns a label absent from the path map & Misrouted Dispatch & LangChain & \href{https://github.com/langchain-ai/langchain/issues/38351}{langchain \#38351} & Fixed \\
5 & Shared router returns targets absent from the edge's path map & Misrouted Dispatch & LangGraph & \href{https://github.com/TauricResearch/TradingAgents/issues/1088}{TradingAgents \#1088} & Fixed \\
6 & Resume reuses a checkpoint from a differently shaped graph & Corrupted Shared State & LangGraph & \href{https://github.com/TauricResearch/TradingAgents/issues/1089}{TradingAgents \#1089} & Fixed \\
7 & Run config drops the thread id on the new context path & Broken Data Handoffs & LangGraph & \href{https://github.com/bytedance/deer-flow/issues/3728}{deer-flow \#3728} & Fixed \\
8 & Tools bind the memory bank at build time, so tenants share it & Corrupted Shared State & LangGraph & \href{https://github.com/vectorize-io/hindsight/issues/2442}{hindsight \#2442} & Fixed \\
9 & Fan-in wired as two edges, not a barrier; fires twice & Corrupted Shared State & LangGraph & \href{https://github.com/MODSetter/SurfSense/issues/1531}{SurfSense \#1531} & Fixed \\
10 & Sub-agent failure counter resets on rewrap; cap unreachable & Stalled \& Runaway Workflows & LangGraph & \href{https://github.com/PurpleAILAB/Decepticon/issues/722}{Decepticon \#722} & Fixed \\
11 & Parallel writes to a channel with no reducer (lost update) & Corrupted Shared State & LangGraph & \href{https://github.com/NVIDIA/NeMo-Agent-Toolkit/issues/2084}{NVIDIA / NeMo \#2084} & Fixed \\
12 & Unconditional edge and router both fire; \icode{END} unreachable & Misrouted Dispatch & LangGraph & \href{https://github.com/NVIDIA/NeMo-Agent-Toolkit/issues/2083}{NVIDIA / NeMo \#2083} & Fixed \\
13 & A final node has no edge to \icode{END}; never terminates & Stalled \& Runaway Workflows & LangGraph & \href{https://github.com/NVIDIA/NeMo-Agent-Toolkit/issues/2082}{NVIDIA / NeMo \#2082} & Fixed \\
14 & Loop's step-count guard is unreachable; iterations unbounded & Stalled \& Runaway Workflows & LangGraph & \href{https://github.com/starpig1129/DATAGEN/issues/29}{DATAGEN \#29} & Fixed \\
15 & Final refinement is returned without passing the reviewer & Role \& Interrupt Violations & Agent Framework & \href{https://github.com/microsoft/OpenAIWorkshop/issues/432}{OpenAIWorkshop \#432} & Fixed \\
16 & Fan-in wired as four edges, not a barrier; fires repeatedly & Corrupted Shared State & LangGraph & \href{https://github.com/NVIDIA/NeMo-Agent-Toolkit/issues/2080}{NVIDIA / NeMo \#2080} & Fixed \\
17 & Router returns a label absent from the path map & Misrouted Dispatch & LangGraph & \href{https://github.com/NVIDIA/NeMo-Agent-Toolkit/issues/2081}{NVIDIA / NeMo \#2081} & Fixed \\
18 & A static edge to \icode{END} contradicts the declared finish node & Stalled \& Runaway Workflows & LangGraph & \href{https://github.com/CopilotKit/CopilotKit/issues/5605}{CopilotKit \#5605} & Fixed \\
19 & Concurrent branches lose a write to a last-value reducer & Corrupted Shared State & LangGraph & \href{https://github.com/minitap-ai/mobile-use/issues/208}{mobile-use \#208} & Fixed \\
20 & Summarizer drops one message per pass; cap is not enforced & Corrupted Shared State & LangGraph & \href{https://github.com/minitap-ai/mobile-use/issues/210}{mobile-use \#210} & Fixed \\
21 & Frontend tool-calls never get a paired tool message & Broken Data Handoffs & LlamaIndex & \href{https://github.com/run-llama/llama_index/issues/22066}{llama\_index \#22066} & Fixed \\
22 & A missing tool-call id is replaced by a fabricated one & Broken Data Handoffs & LlamaIndex & \href{https://github.com/run-llama/llama_index/issues/22068}{llama\_index \#22068} & Fixed \\
23 & Factory shares one initial-state dict across all workflows & Corrupted Shared State & LlamaIndex & \href{https://github.com/run-llama/llama_index/issues/22069}{llama\_index \#22069} & Fixed \\
24 & Parallel branches collide on a shared human-response key & Corrupted Shared State & LlamaIndex & \href{https://github.com/run-llama/llama_index/issues/22070}{llama\_index \#22070} & Fixed \\
25 & An if/else branch left unwired resolves to no target & Misrouted Dispatch & LangGraph & \href{https://github.com/Haohao-end/openagent/issues/248}{openagent \#248} & Fixed \\
26 & Rename handler reads a key the tool callback never emits & Broken Data Handoffs & LangChain & \href{https://github.com/langchain-ai/langchain/issues/38465}{langchain \#38465} & Fixed \\
27 & Message channel has no reducer; one write wipes the history & Corrupted Shared State & LangGraph & \href{https://github.com/starpig1129/DATAGEN/issues/30}{DATAGEN \#30} & Fixed \\
28 & Concurrent runs overwrite a shared client's session token & Corrupted Shared State & LangGraph & \href{https://github.com/moorcheh-ai/memanto/issues/884}{memanto \#884} & Fixed \\
29 & One in-memory store is shared by all concurrent tenants & Corrupted Shared State & LangGraph & \href{https://github.com/Drakkar-Software/OctoBot/issues/3515}{OctoBot \#3515} & Fixed \\
30 & A shared thread-id slot lets an approval resume the wrong run & Corrupted Shared State & LangGraph & \href{https://github.com/Drakkar-Software/OctoBot/issues/3516}{OctoBot \#3516} & Fixed \\
31 & Child sessions share the parent's live skillbook and operations & Corrupted Shared State & PydanticAI & \href{https://github.com/kayba-ai/agentic-context-engine/issues/134}{kayba-ai / ACE \#134} & Fixed \\
32 & A retried task leaves its stale result visible to dependents & Corrupted Shared State & LangGraph & \href{https://github.com/pipeshub-ai/pipeshub-ai/issues/2627}{pipeshub-ai \#2627} & Fixed \\
33 & Failed runs leak entries in a process-wide future registry & Corrupted Shared State & LangGraph & \href{https://github.com/apconw/Aix-DB/issues/222}{Aix-DB \#222} & Fixed \\
34 & Self-loop exits only on an LLM-set flag; no bound of its own & Stalled \& Runaway Workflows & LangGraph & \href{https://github.com/CopilotKit/CopilotKit/issues/5606}{CopilotKit \#5606} & Confirmed \\
35 & A transition targets a node outside its own graph & Misrouted Dispatch & parlant & \href{https://github.com/emcie-co/parlant/issues/819}{parlant \#819} & Confirmed \\
36 & Message channel uses list addition, not the message reducer & Corrupted Shared State & LangGraph & \href{https://github.com/langwatch/langwatch/issues/5159}{langwatch \#5159} & Confirmed \\
\bottomrule
\end{tabular}
\par}
\vspace{0.6em}}]

\appendices

\section{Previously Unknown Bugs: Complete List}
\label{app:wild}

\autoref{tab:wild-full} lists all $36$ previously unknown workflow bugs that \tool surfaced in
deployed open-source projects and that the projects' maintainers have fixed or confirmed, across
$21$ repositories. The Status column reads \emph{Fixed} where a repair has been merged upstream, and
\emph{Confirmed} where a maintainer has acknowledged the bug and no repair has been merged yet.
Its first $15$ rows are those of \autoref{tab:wild} in \autoref{sec:eval:rq3}.

\tcbset{codepanel/.style={enhanced, breakable, colback=gray!4, colframe=archSlate, boxrule=0.6pt,
  arc=2pt, left=4pt, right=4pt, top=3pt, bottom=3pt, toptitle=1.5pt, bottomtitle=1.5pt,
  colbacktitle=archSlate, coltitle=white, fonttitle=\footnotesize\sffamily\bfseries,
  fontupper=\footnotesize, fontlower=\footnotesize, middle=2.5pt,
  segmentation style={archSlate, dashed, line width=0.4pt}}}
\newcommand{\panelhead}[1]{{\footnotesize\sffamily\bfseries\color{archSlate}#1}\par\nobreak\vspace{1pt}}
\lstdefinestyle{panelcode}{language={}, basicstyle=\fontfamily{zi4}\fontsize{6.5}{7.6}\selectfont,
  columns=fullflexible, keepspaces=true, upquote=true, literate={-}{{-{}}}1,
  showstringspaces=false, aboveskip=2pt, belowskip=0pt, xleftmargin=0pt, frame=none}

\section{LLM-as-Judge Prompt}
\label{app:llm-prompt}

Each LLM judge in \autoref{tab:baselines} receives one workflow per call, with this prompt.

\begin{tcolorbox}[codepanel, title={System prompt}]
You are a static analyzer for LLM agent workflows. You will be given the source definition of ONE
workflow---either Python code that builds a graph, or a JSON flow. Decide whether it contains a
bug, considering these six categories of workflow defect:
\begin{enumerate}[label=\arabic*., leftmargin=1.2em, itemsep=0.5pt, topsep=1pt, parsep=0pt]
\item \textbf{Liveness \& Termination}---whether every node can be triggered, loops can terminate,
  and there are no unreachable or dead-end nodes.
\item \textbf{Type \& Data Contracts}---whether the data types/sockets at the two ends of a
  connection are compatible, and whether each required input has a producer.
\item \textbf{Routing \& Resolution}---whether every routing/edge target actually exists,
  conditional branches are correctly gated, and no output is left dangling.
\item \textbf{State \& Concurrency}---whether concurrent writes to a shared channel are merged by a
  reducer, whether a stateful instance is unsafely shared/aliased, and whether reads and writes
  are ordered.
\item \textbf{Composition}---whether a sub-workflow or sub-graph's ports are bound correctly to its
  parent.
\item \textbf{Capability \& Protocol}---whether a node's declared capabilities/tools are compatible
  with its role and protocol, and whether interrupts are ordered correctly.
\end{enumerate}
These six are only the AREAS to consider; you must find any actual defect yourself from the source.

\vspace{1.5pt}
Work ONLY from the source you are given. Do NOT use web search, do NOT look up the original issue
report or its fix, and do NOT rely on any external or memorized knowledge of this particular
workflow---judge solely by analyzing the provided code/flow. (You have no tools and no network
here; this is also enforced structurally.)

\vspace{1.5pt}
For EACH of the six categories, decide whether the workflow has a bug of that category. When yes,
you MUST point to WHERE, using an identifier exactly as it appears in the source: a node name/id,
or an edge written as ``\icode{source->target}''. Do not use line numbers, and do not describe the
location in prose.

\vspace{1.5pt}
Respond with ONLY a JSON object of exactly this shape, nothing else:
\begin{lstlisting}[style=panelcode]
{
  "liveness_termination": {"has_bug": true|false,
      "location": "<node id or src->dst; empty if no bug>"},
  "type_contracts": {"has_bug": true|false, "location": "<...>"},
  "routing_resolution": {"has_bug": true|false, "location": "<...>"},
  "state_concurrency": {"has_bug": true|false, "location": "<...>"},
  "composition": {"has_bug": true|false, "location": "<...>"},
  "capability_protocol": {"has_bug": true|false, "location": "<...>"}
}
\end{lstlisting}
\tcblower
\panelhead{User message}
\begin{lstlisting}[style=panelcode, aboveskip=0pt, breaklines=true, breakatwhitespace=true,
  breakindent=2em]
# This is a <framework> agent workflow (<Python definition | exported flow.json | C# definition>).
# Analyze the workflow below for bugs, working only from this source.
<the workflow's source without comments, docstrings or free text>
\end{lstlisting}
\end{tcolorbox}

\end{document}